\documentclass[pre,twocolumn,superscriptaddress,longbibliography]{revtex4-1}
\usepackage[english]{babel}
\usepackage[usenames,dvipsnames,svgnames,table]{xcolor}
\usepackage{hyperref}
\usepackage{amsmath, amsfonts,amssymb, mathtools}
\hypersetup{colorlinks=true,  linkcolor=NavyBlue, citecolor=PineGreen,urlcolor=cyan}
\usepackage{physics}
\usepackage{graphicx,color}
\usepackage[normalem]{ulem}

\usepackage[utf8]{inputenc}

\usepackage[mathscr]{euscript}

\graphicspath{{./Figures/}}

\renewcommand{\a}{\alpha}

\newcommand{\avg}[1]{\left< #1 \right>}

\usepackage{bm}

\usepackage{array, makecell}
\usepackage{boldline}

\usepackage{pifont}
\newcommand{\cmark}{\ding{51}}%
\newcommand{\xmark}{\ding{55}}%

\begin{document}
	
\title{Finite-size effects in the microscopic critical properties of jammed configurations: A comprehensive study of the effects of different types of disorder}

\author{Patrick Charbonneau}
\affiliation{Department  of  Chemistry,  Duke  University,  Durham,  North  Carolina  27708,  USA}
\affiliation{Department  of  Physics,  Duke  University,  Durham,  North  Carolina  27708,  USA}

\author{Eric I. Corwin}
\affiliation{Department  of  Physics  and  Material  Science  Institute, University  of  Oregon,  Eugene,  Oregon  97403,  USA}

\author{R. Cameron Dennis}
\affiliation{Department  of  Physics  and  Material  Science  Institute, University  of  Oregon,  Eugene,  Oregon  97403,  USA}

\author{Rafael \surname{Díaz Hernández Rojas} }
\affiliation{Dipartimento di Fisica, Sapienza Universit\`a di Roma, Rome, 00185, Italy}

\author{Harukuni Ikeda}
\affiliation{Graduate School of Arts and Sciences, The University of Tokyo 153-8902, Japan}

\author{Giorgio Parisi}
\affiliation{Dipartimento di Fisica, Sapienza Universit\`a di Roma, Rome 00185, Italy}
\affiliation{INFN, Sezione di Roma1, and CNR-Nanotec, unit\`a di Roma, Rome 00185, Italy}

\author{Federico Ricci-Tersenghi}
\affiliation{Dipartimento di Fisica, Sapienza Universit\`a di Roma, Rome 00185, Italy}
\affiliation{INFN, Sezione di Roma1, and CNR-Nanotec, unit\`a di Roma, Rome 00185, Italy}

\begin{abstract}
	Jamming criticality defines a universality class that includes systems as diverse as glasses, colloids, foams, amorphous solids, constraint satisfaction problems, neural networks, etc. A particularly interesting feature of this class is that small interparticle forces ($f$) and gaps ($h$) are distributed according to nontrivial power laws. A recently developed mean-field (MF) theory predicts the characteristic exponents of these distributions in the limit of very high spatial dimension, $d\rightarrow\infty$ and, remarkably, their values seemingly agree with numerical estimates in physically relevant dimensions, $d=2$ and $3$.
	These exponents are further connected through a pair of inequalities derived from stability conditions, and both theoretical predictions and previous numerical investigations suggest that these inequalities are saturated. Systems at the jamming point are thus only marginally stable. Despite the key physical role played by these exponents, their systematic evaluation has yet to be attempted. Here, we carefully test their value by analyzing the finite-size scaling of the distributions of $f$ and $h$ for various particle-based models for jamming. Both dimension and the direction of approach to the jamming point are also considered. We show that, in all models, finite-size effects are much more pronounced in the distribution of $h$ than in that of $f$. We thus conclude that gaps are correlated over considerably longer scales than forces. Additionally, remarkable agreement with MF predictions is obtained in all but one model, namely near-crystalline packings. Our results thus help to better delineate the domain of the jamming universality class. We furthermore uncover a secondary linear regime in the distribution tails of both  $f$ and $h$. This surprisingly robust feature is understood to follow from the (near) isostaticity of our configurations.
\end{abstract}

\maketitle

\section{Introduction}
Jammed systems may lack dynamics, but their study is far from motionless. A surge of physical interest over the past couple of decades has indeed led to marked advances~\cite{puz-book,berthier_biroli-review-2011, liu_nagel_review_2010,mft-review_2010,torquato_review_2010,exact_mft_review, jamming_van_hecke,review_edwards_statmech-jamming}. This sustained interest stems partly from jamming being observed in systems as varied as grains, foams, and emulsions, and partly from jamming exhibiting features encompassed in few universality classes. The mix of ubiquity and universality has motivated the search for a common framework to explain the pervasiveness of jammed systems and their properties, starting with the seminal works of Liu, Nagel and coworkers~\cite{liu_nagel_1998,ohern_liu_2003}. It has since become clear that although different systems reach jamming by tuning different physical variables, several properties \emph{near} and \emph{at} the onset of jamming are shared by all of them. In other words, the same underlying physics should be responsible for the jamming phenomenology. Even though a fully comprehensive theory remains to be formulated, a major step forward has been the discovery that this jamming point is critical and gives rise to a phase transition, albeit an out-of-equilibrium one~\cite{universal_microstructure}.

Attempts to better understand jamming~\cite{berthier_biroli-review-2011} commonly  focus on systems of frictionless spherical particles\cite{torquato_review_2010}, which are central to a fairly wide universality class (see below).
An outstanding example of the theoretical analysis that can be achieved by such geometric simplification is the recently developed mean-field (MF) theory~\cite{exact_mft_1,exact_mft_2,exact_mft_3,exact_mft_review,fractal_fel, puz-book} that describes --exactly, in the infinite-dimensional limit-- the behavior of glass-forming liquids from the point they fall out of equilibrium up to jamming. Even though one might expect this theory only to be valid in high spatial dimensions, near jamming it describes many of the critical properties observed in dimensions as low as $d=2$ and $d=3$ \cite{jamming_criticality_rem_bucks,lerner_low-energy_excitations,puz-book,exact_mft_review}. (A different criticality is observed in quasi-one-dimensional systems~\cite{ikeda2020,moore_marginally_2020}.)
Jamming criticality is peculiar because not only thermodynamic variables, \textit{e.g.}, the pressure or bulk and shear moduli, but also collective quantities, such as the mean square displacement and the average contact number, scale critically with the distance from the jamming point. More specifically, denoting the configuration density (or packing fraction)  $\phi$ and its value at the onset of jamming $\phi_J$, several quantities either jump discontinuously or scale as power laws, $\abs{\phi -\phi_J}^\mu$, as the jamming point is approached~\cite{ohern_liu_2003,liu_nagel_review_2010,jamming-dynamic-criticality}. Although $\phi_J$ depends sensitively on the preparation protocol --thus giving rise to a density continuum of jamming points \cite{hopkins_disordered_2013,md_ls,gardner_pnas_2016,torquato_review_2010,torquatoRobustAlgorithmGenerate2010,jiaoNonuniversalityDensityDisorder2011}-- $\mu$ is often surprisingly independent of dimensionality and polydispersity\cite{ohern_liu_2003}. And even though different interaction potentials may yield different exponents for a given quantity, this dependence can often be trivially accounted for~\cite{liu_nagel_review_2010,jamming-dynamic-criticality,jamming_van_hecke}.
Importantly, once a jammed state is reached for a given potential, the resulting configuration is an equally valid jammed state for any other potential~\cite{jamming_criticality_rem_bucks} .

However broad this universality class may be, it does not prevent $\mu$ from depending on whether the jamming point is approached either from below [\textit{i.e.}, from the under-compressed (UC) phase, $\phi\to\phi_J^-$] or from above [over-compressed (OC) phase, $\phi\to\phi_J^+$]. A salient example is pressure, $P$, which scales as $P \sim \abs{\phi-\phi_J}^{\pm 1}$~\cite{ohern_liu_2003,jamming-transition-temperature}, i.e., $\mu_{\pm} = \pm 1$ as $\phi\to \phi_J^\pm$. In the UC case, pressure thus diverges as density approaches $\phi_J$, as found in granular materials or glass-formers made out of infinitely hard particles~\cite{mft-review_2010}. Conversely, in the OC case, pressure vanishes linearly as the packing fraction is brought down to  $\phi_J$, as found in soft-harmonic particles~\cite{ohern_liu_2003}.
Another important example is the average contact number, $\overline{z}$. Simulations of harmonic soft spheres, for instance, show that $\overline{z}$ exhibits a discontinuity exactly as $\phi\to\phi_J^-$, and then grows as $\overline{z}(\phi)-\overline{z}(\phi_J)\sim(\phi-\phi_J)^{1/2}$ for $\phi>\phi_J$~\cite{ohern_liu_2003}. This discontinuity can be related to the condition that the number of contacts in a configuration should exactly match its number of degrees of freedom, \textit{i.e.}, the onset of isostaticity~\cite{moukarzel_isostatic_1998,liu_nagel_review_2010,jamming_van_hecke}.
Recent studies have further verified the expected finite-size scaling of $P$, $\overline{z}$, and the bulk and shear moduli for a wide variety of potentials in $d=2$ and 3~\cite{liu_size_scaling,liu_jamming_finite_systems}. A Widom-like scaling function has further been derived for these variables as well as for the configurational energy and shear stress~\cite{scaling_ansatz_jamming}. Furthermore, various studies have identified correlation lengths associated to the characteristic length scales of vibrational response to perturbations~\cite{length_scale_rigidity_2013,length_scale_transversal_2013}, the fluctuations in the number of contacts~\cite{hexner_two_length_scales_2018,can_large_packing_2019}, and the fluctuations of particle mobility~\cite{jamming-dynamic-criticality}, all of which diverge at the jamming point.
These observations for  thermodynamic variables and bulk properties provide some of the strongest evidence in support of the critical nature of the jamming transition.

Remarkably, some of the \emph{microscopic} structural properties of jammed configurations, such as the distributions of contact forces and interparticle gaps, are also expected to exhibit nontrivial critical scalings. In particular, in a jammed configuration of $N$ spherical particles with center positions $\{\vb{r}_i\}_{i=1}^N$ and diameters $\{\sigma_i\}_{i=1}^N$, one can define a dimensionless gap between any pair of particles, $h_{ij} = \frac{\abs{\vb{r}_i - \vb{r}_j}}{\sigma_{ij}} -1$, with $\sigma_{ij}=(\sigma_i+\sigma_j)/2$. Because jammed packings are disordered, gap values are randomly distributed, but theoretical predictions~\cite{fractal_fel} state that the distribution of small gaps should scale as
\begin{equation}\label{eq:pdf-gaps}
g(h)\sim h^{-\gamma}, \qquad \text{with } \gamma = 0.41269\dots
\end{equation}
Similarly, the distribution of small contact forces is predicted to scale algebraically, $p(f) \sim f^{\theta}$, but initial reports found a strong dependence of $\theta$ on dimensionality and jamming protocol, in apparent contradiction with the theoretical expectation~\cite{degiuli_pnas_2014}. This paradox was resolved by recognizing that two different types of forces contribute in this regime~\cite{jamming_criticality_rem_bucks,degiuli_pnas_2014}. Opening the contact between a pair of particles can indeed give rise to two distinct responses: (i) a localized rearrangement of neighboring particles or (ii) a displacement field that extends over the whole configuration, without decaying with distance. The former is associated with a buckling motion, and hence remains localized; the latter is associated with a correlation length of the same order as the system size, and hence is a clear example of the criticality of jammed packings. Considering these two types of forces separately yields two power laws with different exponents,
\begin{subequations}\label{eq:pdf-forces}
	\begin{align}
	p(f_\ell) & \sim f_\ell^{\theta_\ell} , & \text{with } & \theta_\ell\simeq 0.17\qc  \label{eq:pdf-f-loc}  \\
	p(f_e) & \sim f_e^{\theta_e} , & \text{with } &  \theta_e = 0.42311\dots \, ; \label{eq:pdf-f-ext}
	\end{align}
\end{subequations}
for localized and extended excitations, respectively. The ability of MF theory~\cite{puz-book,fractal_fel,exact_mft_review} to predict the nontrivial values of $\gamma$ and $\theta_e$ is considered a major analytical success. MF theory, however, does not directly predict $\theta_\ell$, because bucklers are an intrinsically low-dimensional feature~\cite{jamming_criticality_rem_bucks}, and are therefore absent from the $d\to\infty$ description.
The critical exponents of gaps and contact forces are also of utmost importance because they are associated with the mechanical stability of jammed packings. By considering the displacement field that follows opening one of the two types of contacts as well as the ensuing closure of gaps to form stabilizing contacts, a pair of inequalities between $\gamma$, $\theta_\ell$, and $\theta_e$ can be derived~\cite{wyart_marginal_stability,lerner_low-energy_excitations},
\begin{subequations}\label{eq:marginality-bounds}
	\begin{align}
	\gamma & \geq \frac{1-\theta_\ell}{2}\qc \label{eq:gamma_theta_loc}\\
	\gamma & \geq \frac{1}{2+\theta_e} \ .  \label{eq:gamma_theta_ext}
	\end{align}
\end{subequations}
MF theory values as well as numerical simulations indicate that both inequalities are in fact saturated, implying that jammed packings are \emph{marginally} stable~\cite{wyart_marginal_stability,wyart_muller_review_marginal_2015}. This result is consistent with the MF description, which always locates the jamming point within a critical Gardner phase that emerges deep in the glass phase and is characterized by the emergence of marginally stable states~\cite{fractal_fel,exact_mft_2,exact_mft_review,gardner_pnas_2016,gardner_perspective,puz-book}.

The picture that coalesces from putting together the exact MF description with the critical scalings for thermodynamic and other variables, and from considering the robustness of numerical experiments for several dimensions and for different protocols~\cite{universal_microstructure,jamming_criticality_rem_bucks,exact_mft_review,mft-review_2010}, suggests that the jamming transition of spherical particles properly defines a universality class. We now know that this class should encompass a broad range of problems and models beyond spherical particles, including the perceptron~\cite{simplest_jamming,franz_critical_perceptron}, neural networks~\cite{jamming_overparametrization,jamming_neural_nets,jamming_multilayer_supervised}, statistical inference~\cite{inference_glassy_nature}, and the SAT-UNSAT transition in continuous constraint satisfaction problems~\cite{jamming_sat_unsat,krzakala_kurchan_landscape_2007}. Recent works have shown that universality persists even when the interactions are nonanalytic, for instance, due to discontinuous forces~\cite{franz_critical_perceptron,franz_linear_spheres}.

Yet, a careful analysis of the values of $\theta_\ell$, $\theta_e$, and $\gamma$ inferred from numerical simulations has not systematically been carried out. Conducting such an analysis is especially important considering that
packings of slightly polydisperse crystals are reported to exhibit a microstructure characterized by exponents that differ considerably from those of Eqs.~\eqref{eq:pdf-gaps} and \eqref{eq:pdf-forces}~\cite{gardner-crystals,tsekenis_jamming_2020}. Additionally, recent works have shown that many of the salient features of spherical packings depend sensitively on particle shape. For instance, introducing even an infinitesimal amount of asphericity changes the universality class~\cite{ikeda_infinitesimal_2020,brito_universality_2018}, in which the isostatic condition no longer holds. An assessment of the extent of the jamming universality class and an accurate test of its many theoretical predictions are therefore in order~\cite{gardner_perspective}.

In this work we systematically analyze the finite-size scaling of the distributions of interparticle gaps and contact forces. These distributions are one of the fundamental consequences of the presumed nontrivial critical behavior of jammed packings, hence their testing is a key step toward rigorously validating a whole set of critical properties. Although a similar analysis has been carried out for the perceptron~\cite{perceptron_size_scaling} and for the gaps distribution of a two-dimensional binary mixture~\cite{ikeda_infinitesimal_2020}, no systematic result exists for jammed packings of spherical particles nor for amorphous packings with other sources of disorder. Here, in addition to analyzing the most common cases of jammed configurations, \textit{i.e.} $ 2d $ polydisperse and $ 3d $ monodisperse packings, we consider two additional sets of jammed packings: (i) polydisperse spheres in a crystalline FCC structure; and (ii) Mari-Kurchan (MK) hard spheres with random shifts distributed uniformly over space~\cite{mk_definition_2011}. By examining the impact of different sources of disorder, we attempt to define precisely which are the most robust features of jamming criticality, and thus better demarcate its physical universality. The rest of this paper is organized as follows. In Sec.~\ref{sec:methods} we describe the models used and the algorithms employed to produce jammed configurations and extract the relevant structural information, \textit{i.e.} the interparticle gaps, $h$, and contact forces associated with extended, $f_e$, and localized, $f_\ell$, displacement fields. We also explain how finite-size effects in the distributions of these structural variables are considered. In Sec.~\ref{sec:3d-HS} we present a detailed analysis of the finite-size effects in jammed configurations of monodisperse spherical particles in $ 3d $, where we reveal the striking contrast of such effects on the distributions $f_e$ and $h$. Then, in Sec.~\ref{sec:other-systems} we present a similar analysis for the other types of systems considered, finding important differences with the results for $d=3$ spherical systems. We nevertheless argue that most of these differences can be explained from the other scaling corrections described in Sec.~\ref{sec:expected scalings}.
Because theory and previous numerical studies suggest that $f_e$ and $h$ are critically correlated across the whole system, we first consider these two quantities. The distribution of localized forces, $f_\ell$, associated with buckling effects is expected to be independent of system size, hence its analysis is postponed to Sec.~\ref{sec:loc-forces}. A discussion and brief conclusion are given in Sec.~\ref{sec:discussion}.

\section{Numerical methods, models systems, and finite-size scaling}\label{sec:methods}

In this section, we describe the numerical techniques used to produce jammed sphere packings, coming from either the OC or the UC phase. Studying independently these two regimes is useful because --as for other critical points-- there is no reason \textit{a priori} to assume that the scalings from above and below $\phi_J$ are the same. Because each of these two phases is identified with different materials, namely granular matter (from the UC regime) and glasses, foams, and colloids (from the OC phase), this verification is an important test of materials universality. We also describe the other models considered, which are chosen to better appraise the extent of the jamming universality class. The methodology employed to analyze the system-size dependence on the distributions of the microstructural variables, Eqs.~\eqref{eq:pdf-gaps} and \eqref{eq:pdf-forces}, is also detailed.

\subsection{Jammed states from the OC phase}\label{sec:methods-jamming-oc}
We first consider three-dimensional configurations of $N$ spheres of equal diameter, \textit{i.e} $\sigma_{ij} = \sigma\ \forall i,j=1,\dots,N$, in a cubic box under periodic boundary conditions. In a certain sense, this choice is the minimal model with which to produce jammed packings. Lower-dimensionality systems inevitably crystallize unless polydisperse mixtures are used, but ordering can be avoided for monodisperse spheres in $d\geq3$. Sphere positions then serve as the only source of disorder. Given the set of vectors of positions $ \{\vb{r}_i\}_{i=1}^N $, the jamming point starting from the OC phase is obtained for the harmonic contact potential,
\begin{equation}
U\qty( \{\vb{r}_i\}_{i=1}^N ) = \frac{\epsilon}{2} \sum_{i,j}  (\sigma - \abs{\vb{r}_i - \vb{r}_j})^2 \ \Theta \qty( \sigma - \abs{\vb{r}_i - \vb{r}_j}  ) \qc
\end{equation}
where $\epsilon$ is a constant that defines the energy scale) and $\Theta$ is the Heaviside step function. Hence, a pair of particles only interacts if there is an overlap between them. Starting in the OC phase with $\phi>\phi_J$ [$\phi=1.02$ in two dimensions (see below) and $\phi=0.792$ in three dimensions] and a uniformly random distribution of spheres in a square box, a series of energy minimization steps and packing fraction reduction steps are performed until the system has just a single state of self stress, which is where jamming criticality occurs~\cite{ohern_random_2002, wyart_marginal_stability, lerner_low-energy_excitations, hopkins_disordered_2013}. Such a state is characterized for having one contact above isostaticity, \textit{i.e.} when the total number of constraints in a system, $N_c$, matches its number of degrees of freedom, $N_{dof}$.
A single state of self stress is required for critical jamming in order to achieve a finite bulk modulus~\cite{broader_view_jamming_2019,liu_size_scaling}. Put differently, the system density is an additional variable that needs to be fixed, and thus requires one additional contact above isostaticity~\cite{donev_pair_2005}. At a given density the FIRE algorithm, a damped dynamics method, is used to achieve force balance in the configuration~\cite{FIRE}. The energy of the configuration is then calculated and the known scaling relation, $U\propto \left(\phi-\phi_J\right)^2$~\cite{jamming_criticality_rem_bucks}, is used to determine by how much the sphere radii should be uniformly decreased to reduce the system energy by a fixed fraction. After several iterations of this procedure, the packing has precisely $N_c=N_sd-d+1$ contacts where $N_s$ is the number of stable particles and thus $N_{dof}=d (N_s -1)$ corresponds to the number of degrees of freedom in a system under periodic boundary conditions. 
A small fraction of particles, termed rattlers, remain unconstrained at jamming and do not contribute to the overall rigidity of the packing\cite{donev_pair_2005,liu_jamming_finite_systems,jamming_criticality_rem_bucks}, thus $N_s=N-N_r$, with $N_r$ denoting the amount of rattlers in a given configuration. In a $d$-dimensional system, these rattlers can be identified as particles with fewer than $d+1$ contacts. Although $N_r$ changes from one configuration to another, $N_r/N$ always lies within a small range of  $\sim 2-3\% $. Only the total number of particles in the system, $N$, is thus reported. After removing rattlers, the dynamical matrix~\cite{jamming_van_hecke} is used to ensure that the packing is jammed. This algorithm is implemented in the pyCudaPacking software using general purpose graphical processing units and quad-precision calculations~\cite{morse2014geometric, charbonneau2016universal, morse2017echoes}. 
Given that our configurations are not subject to any external force, once the jamming point is reached the $N_c$ dimensional vector of forces magnitudes, $\underline{\bm{f}}$, is obtained as the nonzero solution to the set of linear equations that impose the condition of mechanical equilibrium\cite{jamming_criticality_rem_bucks}:
\begin{equation}\label{eq:forces equilibrium}
\mathcal{S}^T \underline{\vb*{f}} = 0 \, ; \quad \mathcal{S}_{\avg{ij}}^{\alpha k} = (\delta_{jk} - \delta_{ik}) n_{ij}^\a \, .
\end{equation}
In this last equation, $\avg{ij}$ with $i<j$ is the index of a contact, $\vb{n}_{ij}$ is the unit contact vector pointing from particle $i$ to $j$, and $\a=1,\dots,d$ indexes its components. (The single state of self stress that results guarantees that $\underline{\vb*{f}}$ is unique.)
Contributions associated with localized buckling displacements, $f_\ell$, are then separated from those that produce extended excitations, $f_e$, using the fact that (with high probability) bucklers are particles with $z_\ell=d+1$ contacts~\cite{jamming_criticality_rem_bucks}. The set $\{f_\ell\}$ is thus taken as the set of forces applied on particles with $z_\ell$ contacts, while $\{f_e\}$ is its complement.

\subsection{Jammed states from the UC phase}\label{sec:methods-jamming-uc}

For configurations initially in the UC regime, an infinitely hard-sphere potential is used and a combination of molecular dynamics (MD) and linear optimization algorithms are employed to approach $\phi_J$ from below. More precisely, we start from a low-density configuration of particles with random positions and use event-driven MD with a Lubachevsky--Stillinger (MD-LS) growth protocol~\cite{md_ls} to increase the (reduced) pressure up to $P=500$. This first step is performed with a fast compression rate in order to avoid any partial crystallization and is then followed by a second, much slower, growth protocol until $P \gtrsim 10^7$. In this way, the MD-LS protocol compresses a low density fluid into an out-of-equilibrium glass at a very high pressure, while still closely following the (phenomenological) equation of state \cite{mft-review_2010,gardner_pnas_2016,md_ls}.
The high pressure configuration is then used as input for the sequential linear programming (LP) algorithm used in Refs.~\onlinecite{artiaco_baldan_2020,diaz_hernandez_rojas_inferring_2020} to produce jammed packings. At each step, the LP algorithm finds the optimal rearrangement of particles that maximizes their radius, considering a linearized version of the nonoverlapping constraint between any pair of particles. On convergence, this algorithm produces a jammed configuration, because neither particle displacements nor size increases are possible. This approach also allows to easily build the full network of contacts at jamming, because genuine contact forces can be identified, up to a proportionality factor, from the active dual variables associated to the nonoverlapping constraints. As with the OC phase, rattlers are removed and only systems with a single state of self stress are considered. Moreover, it is easy to show that the contact forces thus obtained also satisfy Eq.~\eqref{eq:forces equilibrium}, and therefore our hard-sphere packings are well defined jammed states.

Using either of the two methods to reach jamming we find that all our configurations have a similar final density, $\phi_J \approx 0.64$, which corresponds to inherent structures of systems that are quenched relatively quickly \cite{mft-review_2010,md_ls,liu_nagel_review_2010,puz-book,jamming_van_hecke, ohern_liu_2003,torquato_review_2010,hopkins_disordered_2013, charbonneauMemoryFormationJammed2021,artiaco_baldan_2020,diaz_hernandez_rojas_inferring_2020}. 
(Fluctuations around the average value of $\phi_J$ decrease for larger system sizes, as first reported in Ref.~\cite{ohern_liu_2003}.)
Some remarks about the differences of the two protocols are nevertheless in order. First, note that independently of how a jammed packing is realized, it must be a minimum of the corresponding free energy \cite{fractal_fel}. And indeed, both of our protocols are implemented to perform such minimization, although in markedly different circumstances. For instance, critical jamming occurs in the $T\to 0$ limit when coming from the OC phase, so the free energy is minimized by finding a energetic ground state of the configuration. The FIRE algorithm allows to perform such energy minimization, and by iteratively decompressing the system until overlaps vanish, we guarantee that the final configuration is also valid when $T=0$. For hard spheres, by contrast, only the entropic contribution to the free energy matters, because the interaction energy is necessarily zero and the kinetic contribution is trivial. Correspondingly, our MD-LS+LP method proceeds by maximizing the entropy of the configuration as the free volume per particle vanishes~\cite{frenkelOrderEntropy2015}. 
But it should be mentioned that harmonic\cite{xuRandomClosePacking2005} and logarithmic contact potentials\cite{charbonneauMemoryFormationJammed2021,arceriVibrationalPropertiesHard2020} can also be used to produce jammed packings from the UC phase.
In our case however, the two different protocols we implemented to reach free energy minima are conceived to follow the specific route of the systems we aim to model: (OC) thermal glass formers, soft particles etc., or (UC) grains, rigid particles and other athermal systems.

\subsection{Other models of jammed packings}
We also investigate the jamming point of three other models.

\emph{Polydisperse disks:} Previous studies strongly suggest that the upper critical dimension of the exact MF theory is  $d=2$~\cite{jamming_criticality_rem_bucks,lerner_low-energy_excitations,liu_jamming_finite_systems}. However, as mentioned above, particles of different sizes must then be utilized to inhibit crystallization. An additional source of disorder is thus introduced by extracting particle radii from a log-normal distribution to achieve a polydispersity--defined as the ratio of standard deviation to mean--of 20\%.  This was achieved by generating a Gaussian random number, $R,$ with parameters $\mu =0$ and $\sigma = \sqrt{ \ln \left ( 0.2^2 + 1 \right ) }$ and setting the radii to be $e^R$. (Note that the radii distribution parameters should not be confused with the particle diameter used in monodisperse systems.) These soft harmonic spheres are initially in the OC regime, and thus the FIRE-based algorithm is used to bring configurations to their jamming point via repeated quenching and decompression steps.

\subsubsection{Crystalline polydisperse spheres} 
Removing randomness from particle positions while keeping size polydispersity as the main source of disorder is achieved by generating jammed packings on the sites of a regular face-centered cubic (FCC) lattice. Radii are drawn from a log-normal distribution with a polydispersity of 3\%. These nearly crystalline packings are brought to critical jamming using the quenching and decompressing FIRE-based protocols for soft spheres initially in the OC phase. Although this type of system displays many of the features associated with traditional glasses~\cite{gardner-crystals}, its distributions of forces and gaps often markedly differ from those predicted by MF theory~\cite{tsekenis_jamming_2020,gardner-crystals}. By using a system with a different crystalline symmetry we aim to quantify such discrepancy.

\subsubsection{Monodisperse Mari-Kurchan (MK) spheres}
The MK model is a MF reference given that, by construction, the properties of MK configurations are roughly independent of dimension. Specifically, we consider $d=3$ systems of monodisperse spheres that interact according to a randomly shifted distance, $D(\vb{r}_i, \vb{r}_j) = \abs{ \vb{r}_i - \vb{r}_j + \vb{A}_{ij}}$, where $\vb{A}_{ij}$ is a quenched random vector drawn uniformly from the total system volume. Introducing random shifts, $\vb{A}_{ij}$, suppresses almost completely correlations due to short loops on the interaction graph. Even if $D(\vb{r}_i, \vb{r}_j) = D(\vb{r}_j, \vb{r}_k) = \sigma$ it is very unlikely that $D(\vb{r}_i, \vb{r}_k) \simeq \sigma$. In other words, while for particles interacting via the usual Euclidean distance neighbors of a given particle are likely also neighbors, in the MK model, almost certainly, they are not. Because this property is also the case for systems using the  Euclidean distance in the $d\to \infty $ limit, it is expected that the microscopic structural properties of MK jammed configurations should follow the MF theory predictions closely. Besides, it has already been verified that the MK model exhibits several features of more usual glass formers~\cite{mk_hopping_2014}, that a Gardner transition also occurs deep in the glass phase~\cite{mk_gardner_2015}, and that contact number fluctuations are critically correlated at jamming~\cite{can_large_packing_2019}. Consequently, any deviation from MF predictions observed for this system can safely be attributed to finite-size corrections, which makes the MK model a particularly useful reference to explain the contrasting scaling effects in the distributions of gaps and contact forces (Sec.~\ref{sec:discussion}). For this model, we consider hard sphere configurations initially in the UC phase, and use the MD-LS and LP algorithms to reach their corresponding jamming point, after going through the liquid and glass phases\cite{mk_hopping_2014,mk_gardner_2015}.

\subsection{Expected finite-size scalings}\label{sec:expected scalings}

To ensure that we sampled all the systems of a given type with the same accuracy, $M_N$ independent configurations are produced for a fixed value of $N$, such that data of  $N\times M_N\gtrsim 10^6$ particles is obtained. (Specific values for each system are given below.) Forces and gaps can then be studied across many orders of magnitude, and finite-size corrections can be systematically identified. Because testing for power-law distributions using logarithmic binning of the probability density function (pdf) leads to poor comparisons (due to the loss of resolution when grouping data in a single bin to produce a smooth trend~\cite{newman_power_laws}), the cumulative distribution function (cdf) is considered instead. Note that if a random variable $x$ is distributed according to a pdf of the form $\rho(x) \sim x^\alpha$ for $\alpha > -1$, then its cdf follows $c(x) \sim x^{1+\alpha}$.

When fitting a distribution to empirical data it should be considered that even if $x$ ideally follows such a distribution all the way down to $x\to 0$, finite sampling inevitably leads to deviations. Here, the situation is further complicated by our consideration of marginals of correlated variables. Gaps and forces distributions of finite $N$ configurations are indeed prone to exhibit deviations from their expected form due to both finite sampling and system-wide correlations. Fortunately, introducing a scaling function, as is usually done in the study of critical phenomena~\cite{amit2005field,MC-book}, can account for both effects, and hence the dependence of the cdf on system size can be carefully teased out.

To derive the size scaling of the distributions of $x$, we first note that in a sample of size $N \gg 1$, we can estimate the order of the smallest value observed in the data, $x_\text{min}$, from the probability mass assigned to the extremes of the distribution:
\begin{equation}
\int_0^{x_\text{min}} \rho(x) \dd{x} \sim x_\text{min}^{1+\alpha} \sim \frac{1}{N}.
\end{equation}
In other words, $x_{\text{min}}$ can be estimated from the weight assigned to the extremal value of the empiric cdf, whence it follows that $x_\text{min} \sim N^{-1/(1+\alpha)}$. Note that strictly speaking in this last equation $N$ should be replaced by $N_c$ when analyzing, for instance, the distribution of contact forces. However, given that $N_c \sim d N$ and that we are mostly concerned with the scaling exponent, we can safely neglect the associated proportionality constants. The behavior of the gap distribution is expected to be similar, in that the amount of particles almost in contact should be self-averaging. Next, we follow the traditional path for analyzing size scaling and write the pdf as
\begin{equation}\label{eq:scaling-p}
\rho(x) \sim N^\beta \tilde{\rho}\qty(x N^{\frac1{1+\alpha}} )
\end{equation}
where $\tilde{\rho}$ is the scaling function of the pdf such that $\tilde{\rho}(x) \sim x^\alpha$ for $x \gtrsim 1$. The exponent $\beta$ can be easily determined by requiring that  $\rho(x)$ exhibits no $N$ dependence for a large enough value of $x$, given that if $N^{\frac1{1+\a}}\ x \gg 1$ the data should follow the expected power-law scaling for any $N$. We thus get that $\beta = -\frac{\alpha}{1+\alpha}$, whence the expressions used for the scalings studied in Ref.~\onlinecite{perceptron_size_scaling} are recovered. For the cumulative distributions, repeating the above analysis for $c(x) \sim N^{-\delta} \tilde{c}\qty(x N^{\frac1{1+\alpha}} )$ gives $\tilde{c}(x) \sim x^{1+\alpha}$, and it immediately follows that $\delta=1$, whence the relevant scaling relation is
\begin{equation}\label{eq:scaling-cdf}
c(x) \sim N^{-1}\  \tilde{c} \qty(x N^{\frac{1}{1+\alpha}} )  .
\end{equation}
Using the correct $\alpha$ should remove any dependence on $N$. Data for different system sizes should then be rescaled such that they follow a common curve, $\tilde{c}$. Finding a good collapse of the curves for different $N$ thus indicates that deviations from the expected power laws fall outside the thermodynamic limit, but are not caused by the variables following a different power-law scaling. Additionally, showing that the system size influences the cdf of a given variable strongly evinces that such a variable is correlated across the whole system. Hence, an upper bound to the correlation length can then be estimated.

We want to stress that for microscopic variables of jammed configurations the situation is conceptually different from that of standard critical phenomena, because the systems are already \emph{at} the critical point. We here do not investigate how the distributions of contact forces and gaps converge to their expected distributions as we move away from $\phi_J$, but instead analyze how the system size affects the range over which power-law scalings are followed. 
As a result, most techniques for size scaling analysis [\textit{i.e.}, computing $\gamma(N)$ and $\theta_e(N)$ by isolating the nonsingular contribution of an appropriate scaling function \emph{away} from $\phi_J$ and then extrapolating to $N\to \infty$] are inapplicable.
Equation~\eqref{eq:scaling-cdf} can nevertheless be used to estimate the scaling functions of the cdf of gaps and forces obtained by integrating Eqs.~\eqref{eq:pdf-gaps} and \eqref{eq:pdf-forces}, respectively.

\begin{figure*}
	\includegraphics[width=0.99\linewidth]{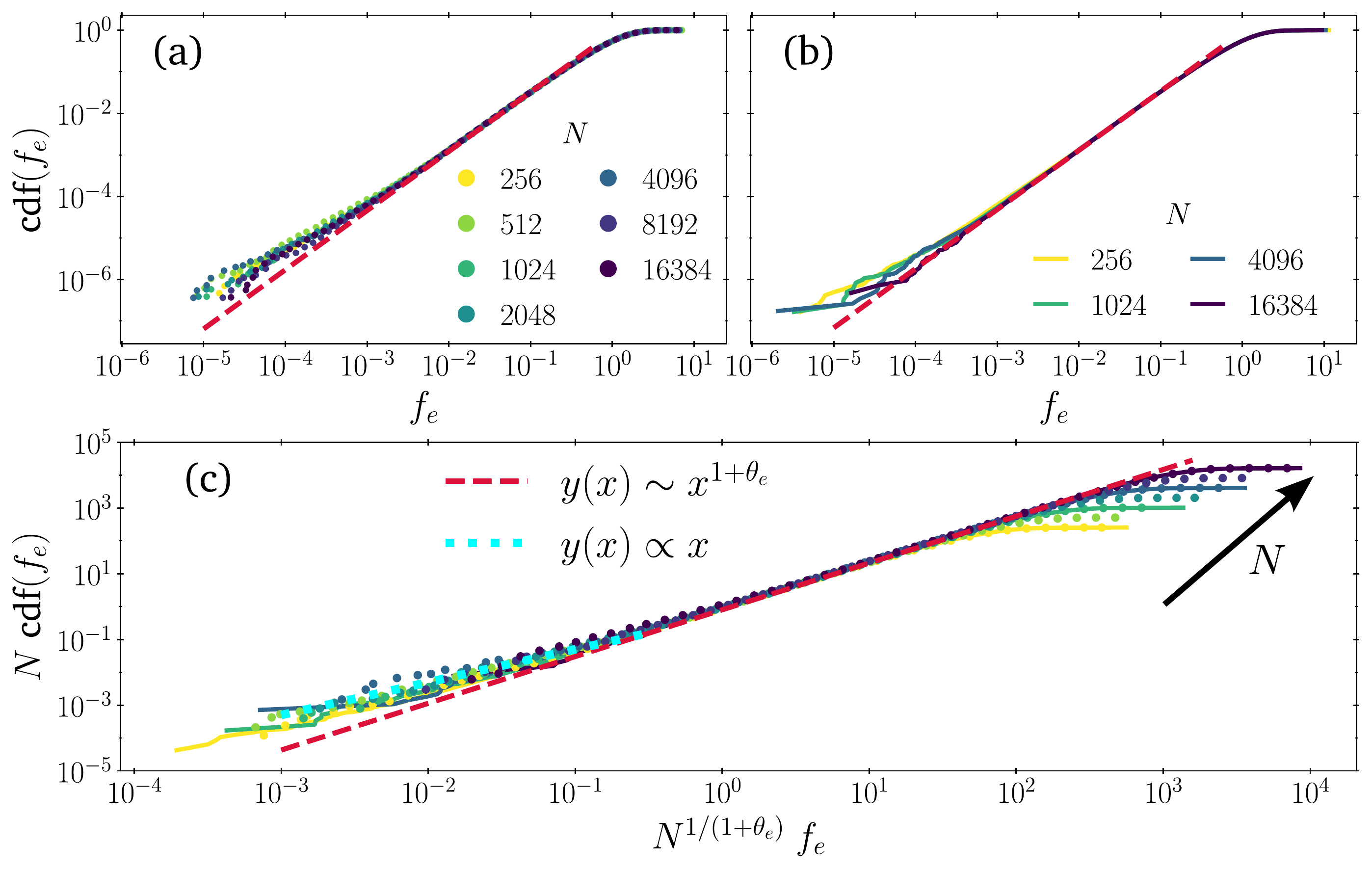}
	\caption{Cumulative distributions of extended contact forces associated with extensive excitations of monodisperse configurations of frictionless spheres for different system sizes $N$, as their jamming point is reached (a) from below (UC) and (b) from above (OC). To better distinguish between the two different regimes, results belonging to the UC (OC) phase are identified by circular markers (solid lines). (c) Rescaling (a) and (b) according to Eq.~\eqref{eq:scaling-cdf} clearly collapses the data. The red dashed line corresponds to the power-law scaling of Eq.~\eqref{eq:gamma_theta_ext}, and shows an excellent agreement between the MF predictions and our numerical results.  The coincidence of results from the UC phase and OC phase for various $N$ confirms that $\theta_e$ is the same when jamming is reached from either direction. In the left tail of the distributions of panel (c) we also include a comparison with the linear scaling (cyan dotted) expected for very small values, following Eq.~\eqref{eq:pdf-f-ext-v2}. When put together, these two behaviors match the predictions given in Eq.~\eqref{eq:general scaling cdf}.
	}
	\label{fig:forces-3d}
\end{figure*}

At the upper critical dimension $d=2$, we expect a logarithmic correction to the size scaling law~\cite{liu_jamming_finite_systems,wang1993monte,ruiz1998logarithmic,Kenna2004},
\begin{equation}
\rho(x)\sim x^\alpha (-\ln x)^{\xi},{\rm for}\ x\ll 1.
\end{equation}
We can then estimate $x_{\rm min}$ as
\begin{equation}
\int_0^{x_{\rm min}} \rho(x)dx \sim x_{\rm min}^{\alpha+1}\left(-\ln x_{\rm min}\right)^{\xi}\sim \frac{1}{N},
\end{equation}
leading to
\begin{equation}
x_{\rm min} \sim
N^{-\frac{1}{1+\alpha}}(-\ln x_{\rm min})^{-\frac{\xi}{1+\alpha}}
\sim N^{-\frac{1}{1+\alpha}}(\ln N)^{-\frac{\xi}{1+\alpha}}.
\end{equation}
Repeating the same argument as above, we get
\begin{equation} \label{eq:scaling-cdf-log-cor}
c(x) \sim N^{-1}(\ln N)^{-\xi}\ \tilde{c}\left(x N^{\frac{1}{1+\alpha}}(\ln N)^{\frac{\xi}{1+\alpha}}\right),
\end{equation}
where the prefactor is chosen such that $c(x)$ does not depend on $N$ for $x\gg x_{\rm min}$. For the cases considered in this work, no theoretical prediction exists for the value of $\xi$, and hence it here serves as a fitting parameter.

\begin{figure*}
	\includegraphics[width=0.99\linewidth]{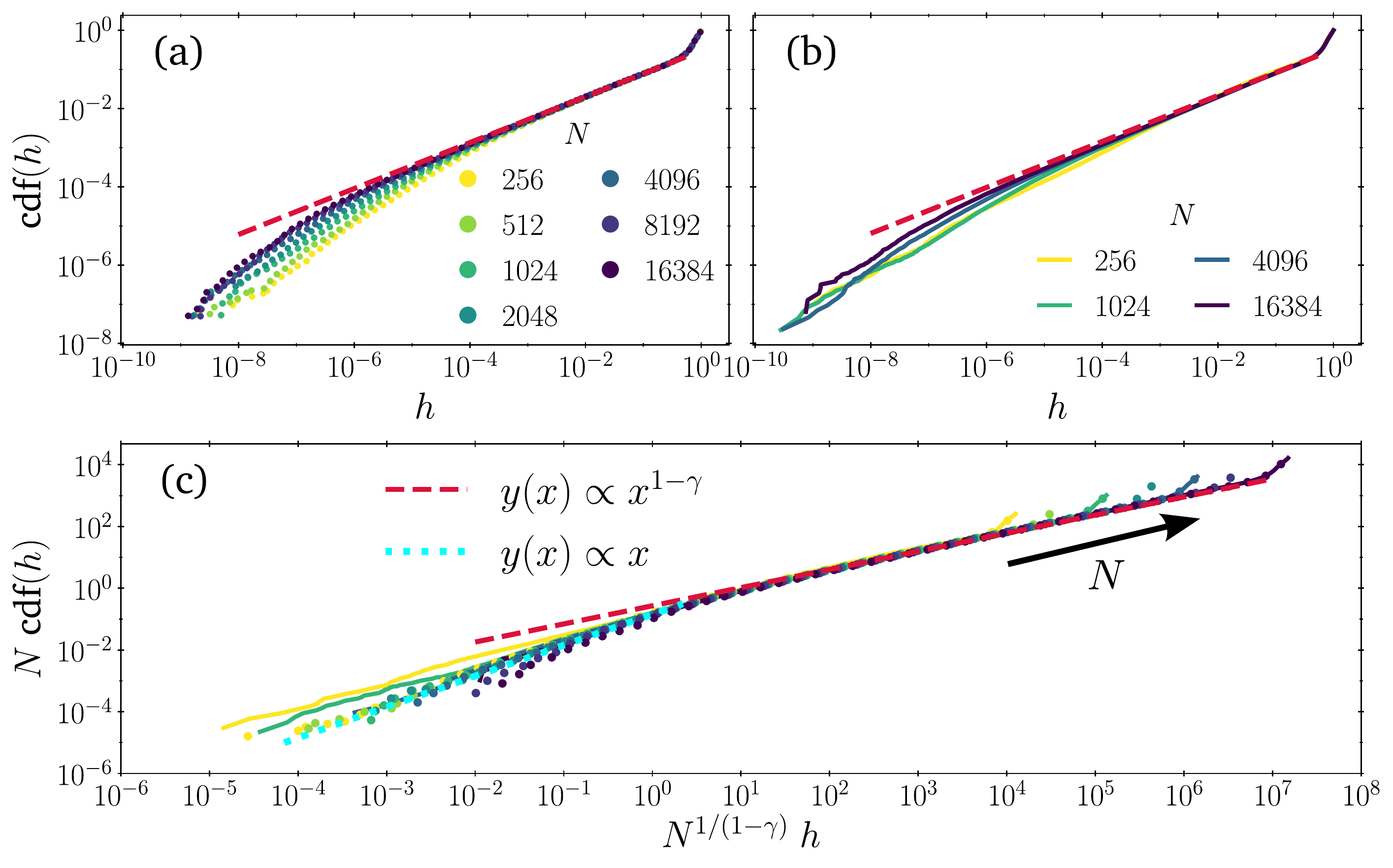}
	\caption{Cumulative distributions of interparticle gaps for the same configurations as in Fig.~\ref{fig:forces-3d}, as their jamming point is reached (a) from below (UC) and (b) from above (OC).  (c) Rescaling (a) and (b) according to Eq.~\eqref{eq:scaling-cdf} shows that finite-size corrections can be accounted for in all cases. For comparison, the power-law scaling derived from MF theory, Eq.~\eqref{eq:pdf-gaps}, is also shown (red dashed line). Once again, the fact that datasets from both phases, \textit{i.e.} UC (markers) and OC (lines), neatly superimpose confirms that the exponents at the jamming point are the same, independently of how  $\phi_J$ is approached. Additionally, the secondary scaling regime $g(h)\sim 1$ of Eq.~\eqref{eq:pdf-gaps-v2}, also predicted by MF theory, can be observed for very small values. Its associated linear cdf is shown (cyan dotted line). These two regimes confirm that the scaling function agrees with our prediction in Eq.~\eqref{eq:general scaling cdf}.
	}
	\label{fig:gaps-3d}
\end{figure*}

We consider yet another correction to Eq.~\eqref{eq:scaling-p} that can also be derived from MF theory. Given that jammed configurations have one extra contact than $N_{dof}$ (see Sec.~\ref{sec:methods-jamming-oc}), the power laws of the microstructural critical variables should be cut off at very small values~\cite{MF_non_spheres,jamming_sat_unsat,ikeda_infinitesimal_2020}. MF theory predicts that interparticle gaps are distributed as $h^{-\gamma}$ only for values larger than a cutoff $h^\star \sim \delta z^\frac{1}{1-\gamma}$, where $\delta z$ is the excess of contacts in a system with respect to isostaticity. In our case, $\delta z \sim 1/N$, so instead of Eq.~\eqref{eq:pdf-gaps} the pdf describing the distribution of $h$ reads
\begin{equation}\label{eq:pdf-gaps-v2}
g(h) \sim \begin{dcases}
N^{\frac{\gamma}{1-\gamma}}\ g_0\qty(h N^{\frac{1}{1-\gamma}}), & h N^{\frac{1}{1-\gamma}} \ll 1 \\
h^{-\gamma}, & h N^{\frac{1}{1-\gamma}} \gtrsim 1
\end{dcases} \, ;
\end{equation}
where $g_0(x)\sim 1$ for $x\ll 1$ \cite{ikeda_infinitesimal_2020}. Analogously, for extended forces Eq.~\eqref{eq:pdf-f-ext} should be replaced by
\begin{equation}\label{eq:pdf-f-ext-v2}
p(f) \sim \begin{dcases}
N^{\frac{-\theta_e}{1+\theta_e}}\ p_0\qty(f N^{\frac{1}{1+\theta_e}}), & f N^{\frac{1}{1+\theta_e}} \ll 1 \\
f^{\theta_e}, & f N^{\frac{1}{1+\theta_e}} \gtrsim 1
\end{dcases} \, ,
\end{equation}
where $p_0(x) \sim 1$ for very small values is to be expected.
Equations~\eqref{eq:pdf-gaps-v2} and \eqref{eq:pdf-f-ext-v2} are indeed consistent with Eq.~\eqref{eq:scaling-p} and, repeating the same arguments as above, it is straightforward to derive that both regimes can be captured by Eq.~\eqref{eq:scaling-cdf} using a single scaling function, such that
\begin{equation}\label{eq:general scaling cdf}
\tilde{c}(x) \sim \begin{dcases}
x\, , & x \ll 1 \\
x^{1+\alpha} \, , & x \gg 1   
\end{dcases} \, .
\end{equation}	
That is, using the correct $\a$ in Eq.~\eqref{eq:scaling-cdf} accounts for size effects that give rise to deviations from the main power-law scaling as well as the appearance of the linear regime in the left tails. By plotting $Nc$ as a function of $N^{\frac1{1+\alpha}} x$ both corrections can thus be tested from a single scaling collapse.

\section{Finite-size effects in $d=3$ systems}\label{sec:3d-HS}

We first consider systems of monodisperse particles in $d=3$ by generating, for each $N$, $M_N$ independent packings, such that $N\times M_N \simeq 2.2\times 10^6$ ($5.5\times 10^6$) particles are considered when the jamming point is approached from the UC (OC) phase. Figure~\ref{fig:forces-3d} shows the distributions of $f_e$ obtained coming from below [UC, Fig.~\ref{fig:forces-3d}(a)] and from above [OC, Fig.~\ref{fig:forces-3d}(b)]. Comparing the results with the theoretical prediction for the power-law scaling reveals an outstanding agreement over at least three decades. More importantly, no visible signature of finite-size corrections can be detected over the range of $N$ considered. To verify more stringently the absence of finite-size effects, we attempted to collapse the different curves by rescaling the extended forces and their cdf following Eq.~\eqref{eq:scaling-cdf}, obtaining the curves reported in Fig.~\ref{fig:forces-3d}(c). This last figure evinces that the same critical distribution of forces is found independently of whether the jamming point is generated from the UC or OC regimes. Yet, it is clear that our packings exhibit an excess of very small forces [an effect more noticeable when jamming is reached from below; see Fig.~\ref{fig:forces-3d}(a)], echoing earlier observations~\cite{jamming_criticality_rem_bucks,lerner_low-energy_excitations,gardner-crystals,perceptron_size_scaling}. Note that the scaling of Eq.~\eqref{eq:scaling-cdf} does not remove these deviations from the predicted distribution. Note also that these deviations roughly occur for the same scaled force, $N^\frac{1}{1+\theta_e} f_e \lesssim 1$. It is therefore likely that forces are subject to size effects caused by the onset of a second power law, $p(f)\sim 1$ [see Eq.~\eqref{eq:pdf-f-ext-v2}]. We get back to this point below.

\begin{figure*}
	\includegraphics[width=\linewidth]{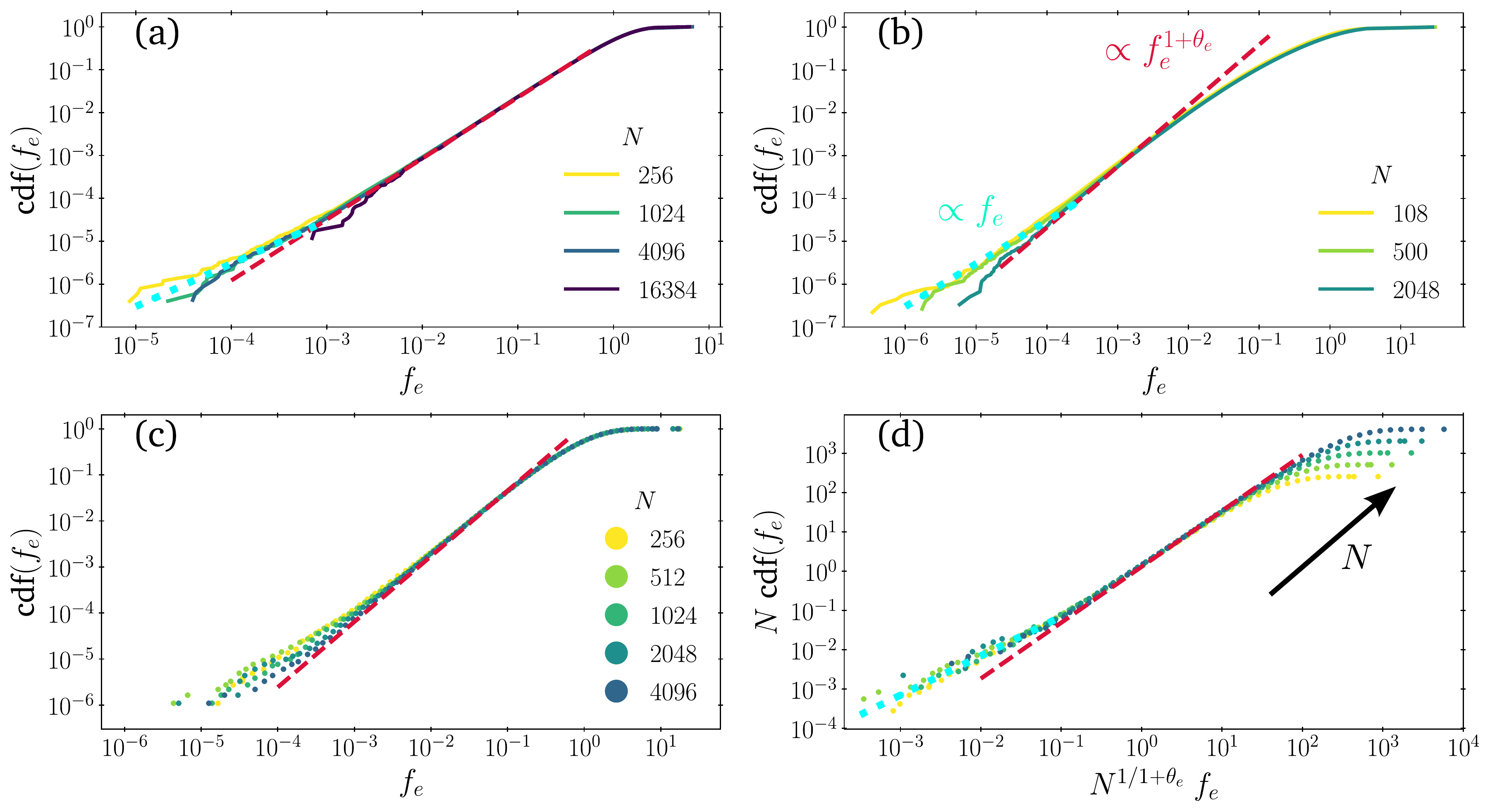}
	\caption{Cumulative distributions of $f_e$ for jammed configurations of (a) $d=2$ polydisperse disks packings, (b) polydisperse spheres with a FCC crystalline structure, and (c) packings using the $d=3$ MK model. Panel (d) depicts the same data from the MK model, rescaled according to Eq.~\eqref{eq:scaling-cdf}; see text for details. Data in the upper (respectively lower) panels were produced as jamming was approached from above (respectively below). The expected power law, Eq.~\eqref{eq:pdf-f-ext} is shown (red dashed lines), as is the secondary linear regime, see Eq.~\eqref{eq:pdf-f-ext-v2} (cyan dotted lines).}
	\label{fig:forces-other-systems}
\end{figure*}

Figure~\ref{fig:gaps-3d}  presents the corresponding cumulative distributions of gaps. The data are also in very good agreement with the predicted scaling of Eq.~\eqref{eq:pdf-gaps}, independently of the direction in which jamming is approached. More importantly, the distributions of $h$ are strongly dependent on system size. In contrast to $p(f_e)$, the scaling correction given in Eq.~\eqref{eq:scaling-cdf} using the MF value of $\gamma$ precisely corrects for such effects over almost \emph{seven} orders of magnitude [Fig.~\ref{fig:gaps-3d}(c)].
The growing deficit of very small gaps as the system size decreases is  another manifestation of the cutoff of the main power law of $g(h)$. It leads to a secondary linear regime, as given in Eq.~\eqref{eq:pdf-gaps-v2}, that is in agreement with the numerical results [Fig.~\ref{fig:gaps-3d}(c)]. This indicates that distances between nearby spheres are significantly modified in finite-size configurations and, consequently, so is the distribution of gaps. This phenomenon is physically interesting. Heuristically, the finite $N$ influence on $g(h)$ can be understood by relying on the marginal stability of jammed packings. In the thermodynamic limit, a system has always enough space to relax any perturbation caused by a contact opening, and hence is always able to reaccommodate particle positions--even if this requires bringing many of them infinitesimally close to each other--in order to guarantee stability.
In a finite system, by contrast, no such unconstrained relaxation can take place. Rearranging an extensive fraction of particles necessarily influences the pair of spheres involved in the contact just opened. There is therefore a certain scale, below which the occurrence of small gaps is disfavored. If the system were further relaxed, then at least one extra contact would form.

At this point, we wish to stress that our results demonstrate the existence of two different types of finite-size corrections to the distributions of extended forces and gaps. The first is a consequence of large-scale correlations and can thus be readily taken into account by the scaling of the cdf given in Eq.~\eqref{eq:scaling-cdf}. Although this correction is practically absent in the forces distribution, for $g(h)$ it is the main source of deviation from the theoretical prediction. The second is a consequence of the critical scalings of Eqs.~\eqref{eq:pdf-gaps} and \eqref{eq:pdf-f-ext} being cut off at very small values. This effect, which is very likely related to the excess contact with respect to $N_{dof}$ (see Sec.~\ref{sec:expected scalings}), affects both microstructural variables and can also be teased out reasonably well using the scaling advanced in Eq.~\eqref{eq:scaling-cdf}. 
We get back to this point in Sec.~\ref{sec:other-systems}, after having considered its signature in other models.

Before concluding this section, it is worth emphasizing that our numerical results are in excellent agreement with the MF, $d\to\infty$ predictions for the power-law scaling of the distributions of both the extended forces and the interparticle gaps. These results  confirm that the jamming criticality of these microstructural variables is robust with respect to changes in the systems dimensionality, all the way down to $d=3$, in agreement with earlier albeit less accurate studies~\cite{universal_microstructure,lerner_low-energy_excitations,degiuli_pnas_2014}. Because results from both OC and UC phases superimpose onto each other, we further conclude that the critical behavior is controlled by the same exponents on both sides of the jamming point.

\section{Finite-size effects in other disordered systems}\label{sec:other-systems}

We next consider the finite-size scaling of the force and gap distributions at jamming for the three other models mentioned above: (i) polydisperse disks, (ii) crystalline polydisperse spheres, and (iii) monodisperse MK spheres. From Sec.~\ref{sec:3d-HS}, we understand that the direction of approach to the jamming point does not influence on the criticality of microstructural variables, so only one such direction is considered for each mode.  The first two approach the jamming point from the OC phase with  $N\times M_N \simeq 5\times10^6$ particles, and the third from the UC phase with $N\times M_N \simeq 10^6$.

\begin{figure}
	\includegraphics[width=0.99\linewidth, height=19cm]{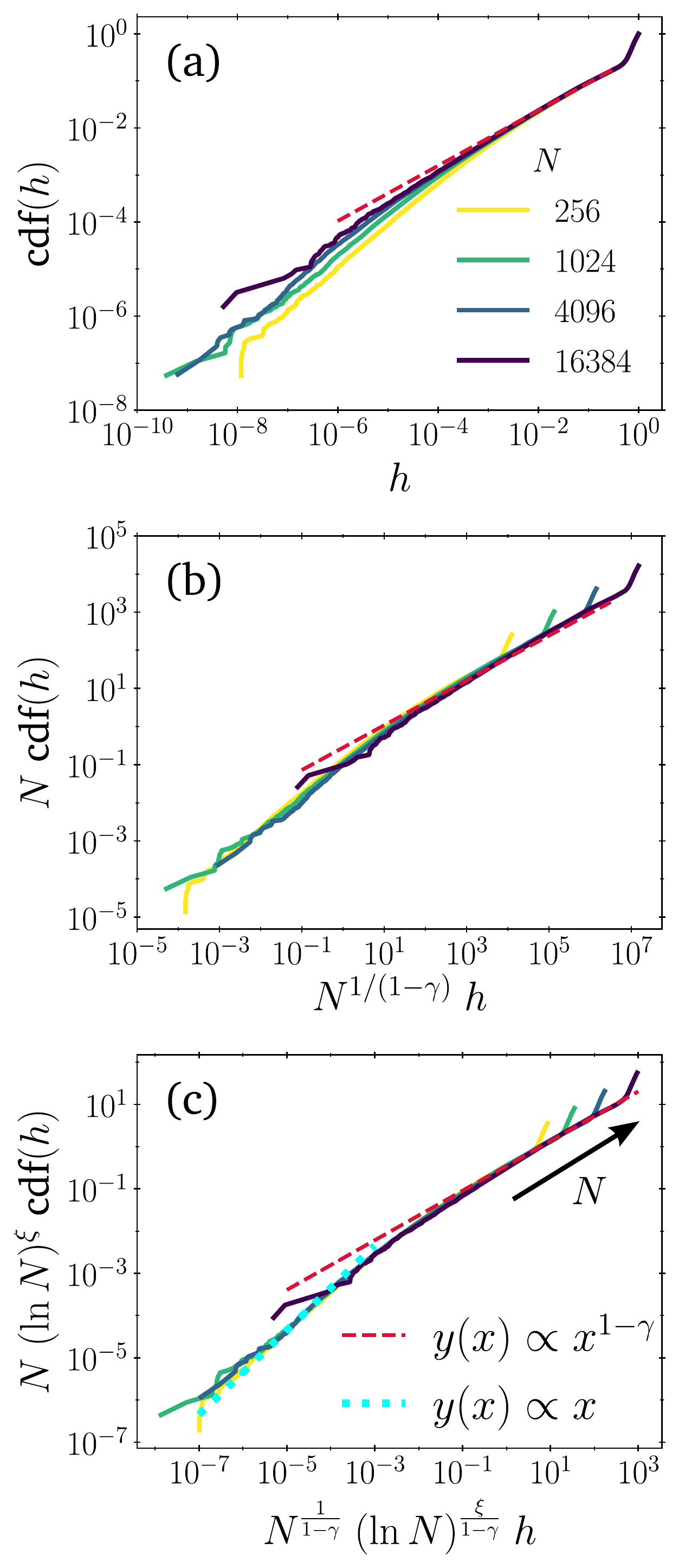}
	\caption{(a) Cumulative distributions of $h$ of jammed configurations of $d=2$ polydisperse disks and different size $N$. (b) Scaling of the different curves following Eq.~\eqref{eq:scaling-cdf} using the MF value of $\gamma$. (c) Same scaling as in (b) but including a logarithmic correction as in Eq.~\eqref{eq:scaling-cdf-log-cor}. Choosing $\xi=-2.5$ then best collapses the results. For reference, the expected power-law scaling is shown  (red dashed line), as is the linear regime given by Eq.~\eqref{eq:pdf-gaps-v2} at very small arguments (cyan dotted line).
	}
	\label{fig:gaps-2d}
\end{figure}

Despite the marked differences between the three models, their distributions of $f_e$ all follow the MF predictions very closely (Fig.~\ref{fig:forces-other-systems}). In Fig.~\ref{fig:forces-other-systems}(a), the $d=2$ packings show a very good agreement with the cdf derived from Eq.~\eqref{eq:pdf-f-ext} over most of the accessible range. In Fig.~\ref{fig:forces-other-systems}(b), results for the FCC symmetry also follow the expected scaling, but because its onset takes place at smaller forces, the range of consistency with the MF power-law scaling is correspondingly reduced.
In Fig.~\ref{fig:forces-other-systems}(c), jammed configurations produced using the MK model exhibit a noticeable, albeit small, dependence on $N$, but this dependence can be removed by rescaling the cdfs according to Eq.~\eqref{eq:scaling-cdf} using the MF value of $\theta_e$ [see Fig.~\ref{fig:forces-other-systems}(d)].
Interestingly, all three systems display an excess of very small contact forces for $f_e \lesssim 10^{-4}$, similarly to what was found for $d=3$ configurations (see Sec.~\ref{sec:3d-HS}). Our results suggest that this effect is due to a crossover to a second regime, in which forces are distributed uniformly, as given by Eq.~\eqref{eq:pdf-f-ext-v2}. A comparison with the corresponding linear behavior in each panel of Fig.~\ref{fig:forces-other-systems} presents a reasonably good agreement, in support of this hypothesis. A more careful analysis would nonetheless be needed to single out the true form of the left tails of $p(f_e)$.

We next consider the finite-size effects on the distribution of gaps of these three systems. From the spacing between different curves in $d=2$ packings, it is clear that such effects are pronounced [Fig.~\ref{fig:gaps-2d}(a)]. Rescaling these distributions following Eq.~\eqref{eq:scaling-cdf} with MF value for $\gamma$ yields a collapse [Fig.~\ref{fig:gaps-2d}(b)] that is not as good as for their $d=3$ counterparts. Section~\ref{sec:expected scalings} anticipated this discrepancy on the basis that $d=2$ is the upper critical dimension for jamming~\cite{liu_jamming_finite_systems}, and hence a logarithmic correction should be included, as in Eq.~\eqref{eq:scaling-cdf-log-cor}. As shown in Fig.~\ref{fig:gaps-2d}(c), with such  correction the data can be robustly collapsed using the MF value of $\gamma$.

\begin{figure}
	\includegraphics[width=0.99\linewidth]{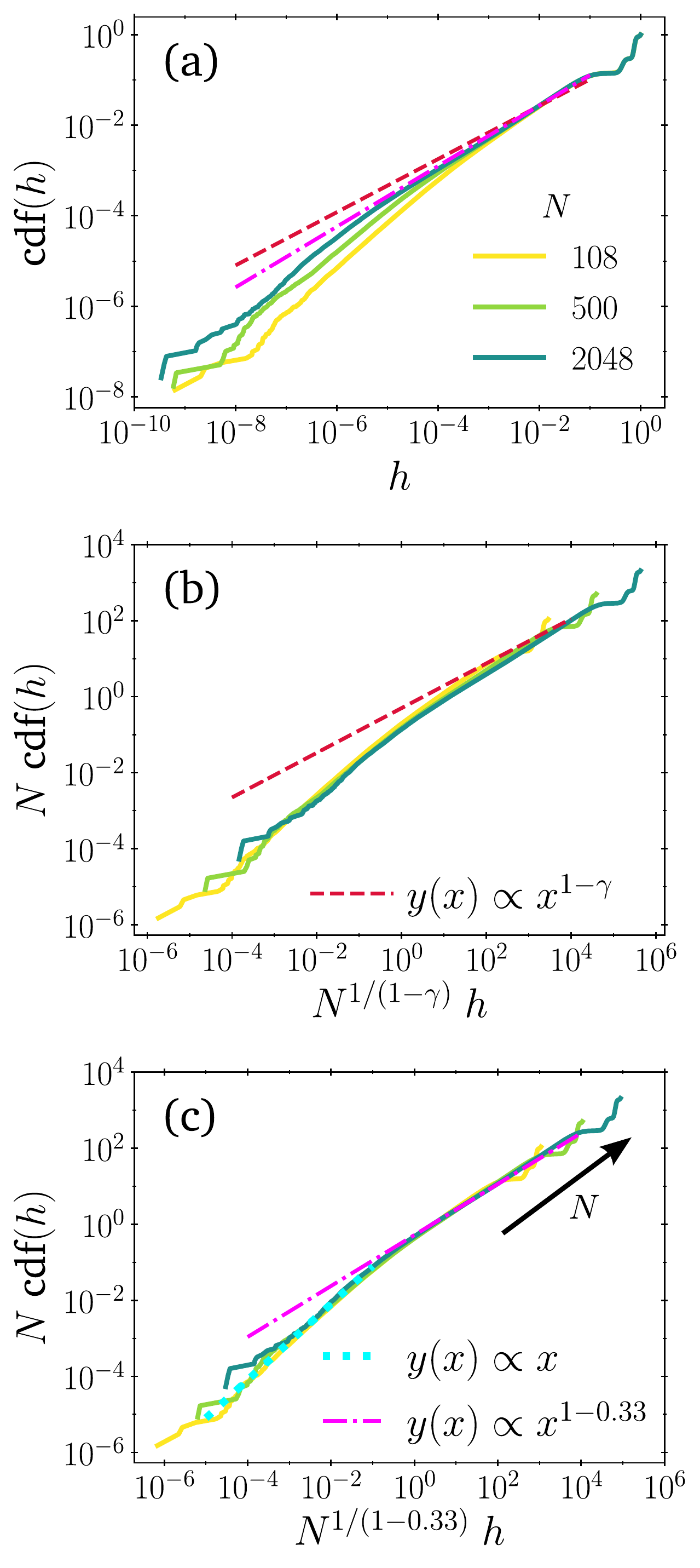}
	\caption{(a) Cumulative distributions of $h$ for jammed configurations of polydisperse spheres with an FCC structure and different $N$. Scaling the different curves according to Eq.~\eqref{eq:scaling-cdf} using (b) the MF value of $\gamma$ and (c) $\gamma_{FCC} = 0.33$. For a clearer comparison, the trend for the expected power-law exponent  (red dashed line) and for $\gamma_{FCC}$ (pink dashed-dotted curve) are shown.  For FCC configurations, unlike for $d=2$ systems, the collapse obtained with the MF value of $\gamma$ is poor over the whole interval considered of the scaled variables [see Fig.~\ref{fig:gaps-2d}(b)]. Note that when  $\gamma_{FCC}$ is used, a linear scaling at very small arguments is recovered (cyan dotted line).}
	\label{fig:gaps-fcc}
\end{figure}

By contrast, gaps distributions in the FCC jammed configurations are best described by a completely different exponent. Figure~\ref{fig:gaps-fcc} clearly shows that [Fig.~\ref{fig:gaps-fcc}(a)]  finite-size corrections are important, but that [Fig.~\ref{fig:gaps-fcc}(b)] a poor collapse is obtained when curves are rescaled using Eq.~\eqref{eq:scaling-cdf} with the MF value of $\gamma$. Using [Fig.~\ref{fig:gaps-fcc}(c)] a different $\gamma_{FCC}\simeq 0.33$, however, satisfactorily captures the $N$ dependence. This confirms previous reports that $\gamma$ is changed in presence of an underlying crystalline structure~\cite{gardner-crystals,tsekenis_jamming_2020}. Reference~\cite{tsekenis_jamming_2020} even found that $\gamma$ depends on the system polydispersity, through the variance of the particle sizes. It is important to stress that finding a smaller $\gamma$ is not merely a matter of scrupulous curve fitting. It also positively violates the marginal stability relations, Eqs.~\eqref{eq:marginality-bounds}, and thus indicate that near-crystals belong to a different universality class than standard amorphous packings of spheres. We comment further on this point in Sec.~\ref{sec:discussion}.

\begin{figure}
	\includegraphics[width=0.99\linewidth]{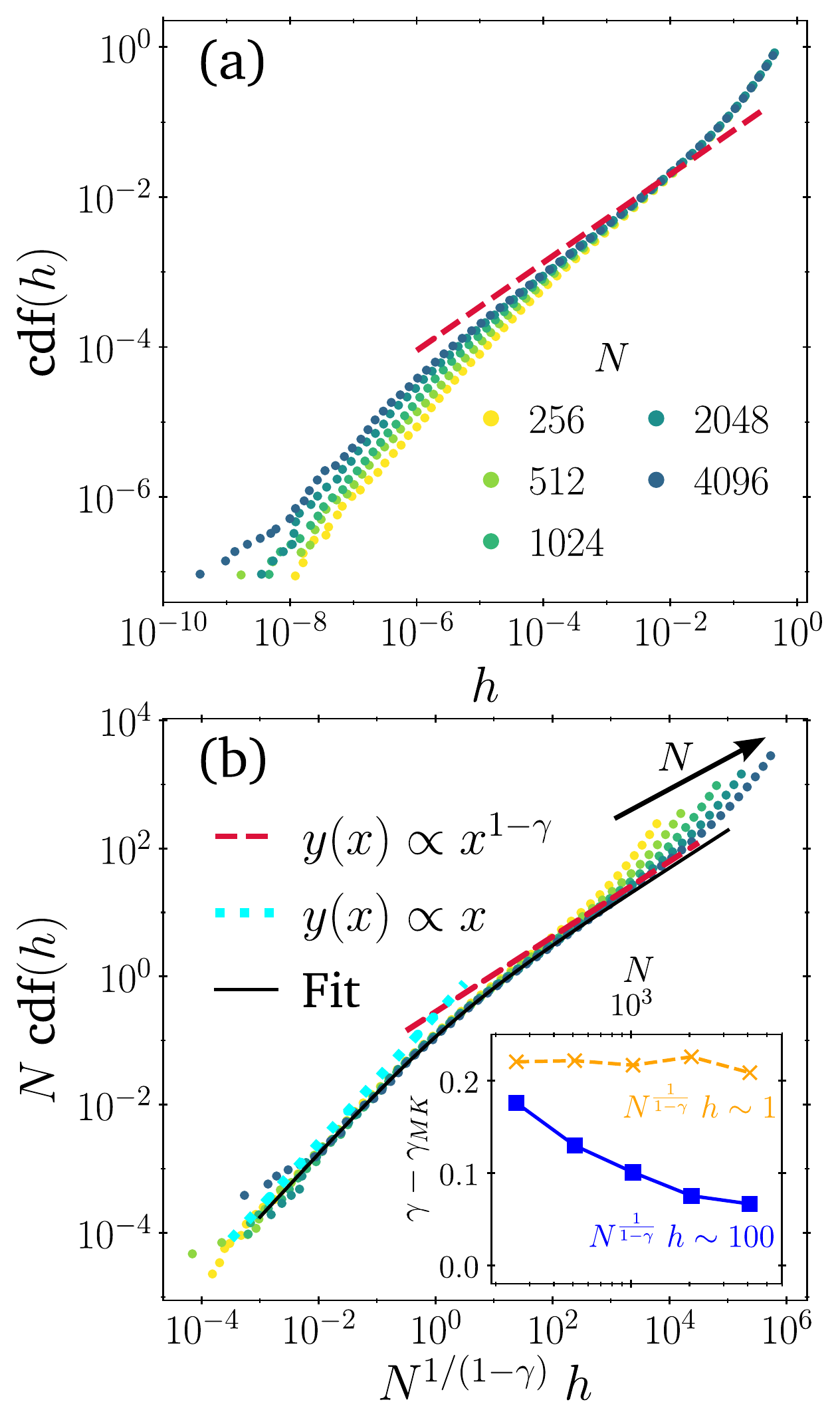}
	\caption{(a) Cumulative distributions of $h$ for jammed configurations of $d=3$ MK systems of different size $N$. (b) Same data but collapsed using the scaling in Eq.~\eqref{eq:scaling-cdf}. Such scaling indicates that $\gamma_{MK} =\gamma $, in agreement with MF theory, although finite-size corrections are particularly important for this model (see main text for discussion). The MF (red dashed lined) and linear (cyan dotted line) behaviors are indicated, as well as the fitting function (solid black) based on Eq.~\eqref{eq:general scaling cdf}, as discussed in the text. Inset: Difference between the MF $\gamma$ and the local slope estimate at two different values of the scaling variable, 1 (crosses, dashed) and 100 (squares, solid). These results suggest that systems orders of magnitude larger would be  needed to recover the pure MF power law (see main text for details).
	}
	\label{fig:gaps-MK}
\end{figure}

\begin{figure*}
	\includegraphics[width=0.99\linewidth]{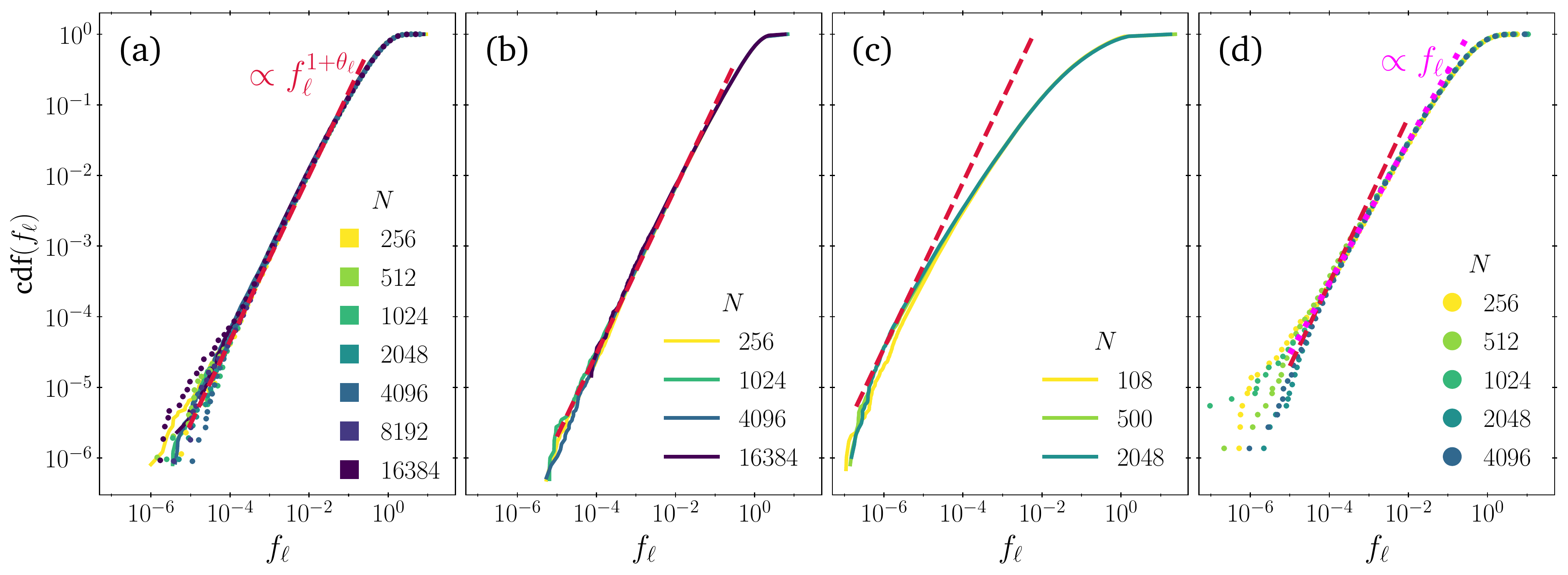}
	\caption{Cumulative distributions of $f_\ell$ for jammed packings of (a) $d=3$ monodisperse spheres, (b) $d=2$ polydisperse disks, (c) polydisperse spheres with FCC structure, and (d) $d=3$ MK model. Solid lines (circular markers) denote data obtained from configurations from the OC (UC) phase. For reference, the expected power law, cdf$(f) \sim f^{1+\theta_\ell}$, with $\theta_\ell = 0.17$, is shown (red dashed lines), and in panel (d) the power-law fit found by inspection for the MK model, cdf$(f) \sim f$, i.e., $\theta_{\ell,MK} = 0$, is also shown (pink dotted line). See text for more details.}
	\label{fig:cdf-fs-localized}
\end{figure*}

Figure~\ref{fig:gaps-MK} presents the gap distributions for the MK model. Here, again, finite-size corrections to $g(h)$ are significant, but now taking the MF value of $\gamma$ in Eq.~\eqref{eq:scaling-cdf} yields a very good collapse, as expected from the MF nature of the model. It is important to note that although individual distributions of $h$ suggest that a smaller exponent would better fit the curves in Fig.~\ref{fig:gaps-MK}(a), doing so worsens significantly the quality of the scaling collapse. This situation is typical of many critical scalings in finite-$N$ systems~\cite{amit2005field,MC-book}. The most reliable way to determine critical exponents remains the finite-size scaling analysis.
It is however surprising that the individual distributions in the MK model, which by construction should be closer to the MF solution, do not display the right gap exponent. Indeed, we observe from Fig.~\ref{fig:gaps-MK}(b) that the scaling variable using the MF value of $\gamma$ is the correct one [data do collapse when plotted versus $\tilde{h}=N^{1/(1-\gamma)} h$], but the slope of the curves in the range covered in our simulations ($10^{-3}< \tilde{h} < 10^3$) is not that predicted by MF theory. An important concern is thus whether this deviation is due to finite-size corrections or whether it indicates a failure of the MK model. In order to resolve the matter, we used the expected form of the scaling function, Eq.~\eqref{eq:general scaling cdf}, to construct a fitting function, $F(\tilde{h})$, that \emph{assumes} the correct behavior of the scaling function for large values of $\tilde{h}$; more specifically, $F(\tilde{h}) = \qty[ (a\tilde{h})^d + (b \tilde{h}^{1-\gamma})^d ]^{1/d}$. The fitting function hence only depends on three parameters and fulfils the condition that $F(\tilde{h})\propto \tilde{h}$ for $\tilde{h}\ll 1$, while the MF form, $\tilde{h}^{1-\gamma}$, is recovered for large values of the scaling variable. Fitting $F(\tilde{h})$ to the largest system size results gives the black line in Fig.~\ref{fig:gaps-MK}(b), which clearly interpolates nicely between both regimes. Therefore, the hypothesis that results for larger MK systems would eventually follow the MF power-law cannot be confuted. The convergence of the slope of the scaling function to the predicted value is nevertheless extremely slow, especially relative to that of other models [see, \textit{e.g.}, Figs.~\ref{fig:gaps-3d}(c) and \ref{fig:gaps-fcc}(c)] or to the distribution of forces in this same MK model [see Fig.~\ref{fig:forces-other-systems}(d)]. One must thus reach very large values of the scaling variable in order to measure the right slope. More precisely, in the inset of Fig.~\ref{fig:gaps-MK}(b) we report the difference of $\gamma$ and our estimation of \emph{local} $\gamma_{MK}$ from the local slope as a function of $N$. Around $\tilde{h}=1$, the slope clearly differs from the MF prediction, but even when $\tilde{h}\sim 10^2$ very large system sizes are needed for it to approach the theoretical exponent. This deviation results in an \emph{apparent} size dependence of the global exponent, \textit{i.e.} $\gamma_{MK} = \gamma_{MK}(N)$, that is substantially more pronounced than for other models at similar $N$.
Such discrepancy likely results from the MK system being fully connected. In contrast with their sparse counterparts, fully connected models indeed require much larger system sizes for thermodynamic power-law scalings to be visible~\cite{lucibello_anomalous_2014,lucibello_finite-size_2014,ferrari_finite-size_2013}.
This feature can be physically understood by recalling that the introduction of random shifts results in neighbors of a given particle (very likely) not being neighbors themselves. A particle can thus have many more contacts than normally allowed in Euclidean space. For instance, it is not uncommon ($\sim1\%$) for particles at jamming to have as many as 12 contacts (the $d=3$ kissing number) or more. In general, particles are thus surrounded by many more particles--both actual and near contacts--than usual hard spheres.
Additionally, jamming densities in this model are much higher than can be achieved with hard spheres. Using our MD-LS+LP algorithm, as well as planting~\cite{mk_gardner_2015} to speed up the growing protocol, results in jamming packing fractions  $\phi_{J,MK} \gtrsim 3.1 $ (cf. $\phi_{J,3d}\simeq 0.64$).
Now, given that $\phi\sim \sigma^{1/d}$, our MK configurations are made out of particles nearly twice as big as those of standard hard spheres. The combination of these two effects is that particles in MK packings are surrounded by a cluster of many relatively large neighbors. The effective size of the system being drastically reduced, finite-size corrections are correspondingly more pronounced. We thus conclude that gaps in the MK model will probably follow the MF power-law scaling, as expected, but only at system sizes orders of magnitude larger than those considered here. In practice the finite-size effects are so important in the distribution of gaps in the MK model that its MF nature is, perhaps paradoxically, a strong limitation to study its MF behavior.

Looking at the whole set of gap distributions, an interesting feature is the robust emergence of a regime of uniform distribution at very small gaps, in a way entirely analogous to the distributions of extended contact forces. We argued in Sec.~\ref{sec:expected scalings} that this truncation of the leading power-law scaling in the distributions likely follows from the combined effect of the additional state of self stress and the system sizes being finite. All the models consistently exhibit this behavior and show very good agreement with the associated linear scaling [see Figs.~\ref{fig:gaps-3d}(c), \ref{fig:gaps-2d}(c), \ref{fig:gaps-fcc}(c) and \ref{fig:gaps-MK}(b)].
The invariance of this secondary power-law scaling with dimensionality, inherent order or other system properties is reassuring, albeit somewhat surprising, given that the leading power-law scaling is more strongly affected by these same effects. The universality of this secondary scaling has been previously predicted~\cite{jamming_sat_unsat} for all models that can be mapped to jamming of spherical particles, and it has been shown to occur even for nonspherical particles~\cite{ikeda_infinitesimal_2020}, provided that their jammed states remain sufficiently close to isostaticity. Such robustness can be understood in part by considering that isostaticity is a global property of the system related to a matching between constraints and degrees of freedom, and not to the specific distributions of its microstructural variables.
Because we have restricted our analysis to packings with exactly $N_c = N_{dof}+1$, the ubiquity of the linear left tails in our distributions supports the hypothesis that the form of $g_0(x)$ [Eq.~\eqref{eq:pdf-gaps-v2}] and $p_0(x)$ [Eq.~\eqref{eq:pdf-f-ext-v2}] is determined by the single state of self stress alone, and not by the inherent structure. It is then remarkable that the same size scaling also seems to capture the behavior of the extremal part of the distributions of gaps and of extended forces, albeit not as evidently for the latter. Our findings are therefore in agreement with Eq.~\eqref{eq:general scaling cdf}.

\section{Cumulative distributions of $f_{\ell}$}\label{sec:loc-forces}

The last microstructural variable we consider is the set of localized forces. Figure~\ref{fig:cdf-fs-localized} presents the probability distributions for all our results.
As expected, this quantity exhibits no clear finite $N$ signature for any of the models, even though some dispersion around the expected behavior is observed in the left tails of $d=3$ monodisperse and MK configurations, [Figs.~\ref{fig:cdf-fs-localized}(a) and \ref{fig:cdf-fs-localized}(d), respectively]. This behavior is expected because the set $\{f_\ell\}$ corresponds to contact forces acting on bucklers, for which opening a weak contact mostly results in localized displacement field~\cite{jamming_criticality_rem_bucks,lerner_low-energy_excitations}. Because opening any of the contacts associated with a buckler only has a non-negligible effect over a few particle layers away from its origin, it is reasonable to assume that their properties should be insensitive to $N$, or to any border or periodic effects.

An intriguing finding is that only the cdf of $d=3$ monodisperse and $d=2$ polydisperse particles follow the known value of $\theta_\ell\simeq0.17$  [see Figs.~\ref{fig:cdf-fs-localized}(a) and \ref{fig:cdf-fs-localized}(d)]. By contrast, FCC structures give rise to no obvious power-law scaling. The FCC arrangement induces strong spatial correlations that seem to suppress the appearance of localized forces, as seen from the smaller slope of the cdf. Observing a distribution with an exponent different from $\theta_\ell$, or actually failing to scale as a power law, is in striking contrast with many other models, and even other crystalline structures~\cite{gardner-crystals}. It nonetheless echoes very recent reports of a dependence of $\theta_\ell$ on geometry for other near-crystals~\cite{tsekenis_jamming_2020}. These considerations highlight the need for further assessment of which aspects of jamming criticality are indeed universal, which are more generically conserved~\cite{ikeda_jamming_crystals}, and which disappear in the presence of long-range spatial constraints.

Although a power-law scaling is also obtained for MK configurations, the best fit to the data is achieved with a unit slope, i.e., $\theta_{\ell,MK}=0$ [see Fig.~\ref{fig:cdf-fs-localized}(d)]. Localized forces are thus distributed uniformly in this model.
A careful analysis suggests that this unexpected distribution is in tune with the spatial properties of MK packings. First, note that even though bucklers follow a different pdf, selecting particles with $z_\ell=d+1$ contacts is still a valid selection criterion. (If their contribution had not been isolated, then the remaining forces would not follow the MF power-law scaling given in Eq.~\eqref{eq:pdf-f-ext}, as it does in Fig.~\ref{fig:forces-other-systems}(c), whereas if both kinds of forces are considered together, their joint pdf  scales as $\approx 1.1$, which differs from the analogous quantity for standard hard spheres~\cite{jamming_criticality_rem_bucks}.) Second, analyzing the distribution of dot products between contact vectors as in Ref.~\onlinecite{diaz_hernandez_rojas_inferring_2020} reveals that particles with $z_\ell$ contacts in MK packings have a very similar distribution as those in standard hard sphere packings. Bucklers thus mainly give rise to a localized response thanks to them having three nearly coplanar contacts and one nearly orthogonal force. In order to understand why localized forces are uniformly distributed, we follow Ref.~\onlinecite{lerner_low-energy_excitations}, which showed that the two types of contact forces are related to two types of floppy modes: extended forces are related to floppy modes that can couple strongly to external perturbations, and hence their response is bulk dominated; and buckling forces are associated to floppy modes of a rapidly decaying displacement field. (The value of $\theta_\ell \approx 0.17$ was estimated from the statistics of \emph{displacements} in the latter.) There is therefore a strong connection between the distribution of forces in bucklers and the particle displacements their floppy modes produce. Now, let us assume that in an MK packing we open a buckling contact, $\avg{ij}$, between particles $i$ and $j$, in order to describe the associated displacement field. In particular, let us focus on the remaining contacts of any of these particles, say $i$. Because of the random shifts, the other particles touching $i$ are (very likely) not constrained by each other nor by the other particles near $i$. Instead, the displacement of each neighbor of $i$ is limited by its own contacts, which are not neighbors themselves, and are typically far apart. By the same token, the effect on the rest of particles in contact with $j$ is determined by secondary contacts that--with high probability--are distant from each other and from $\avg{ij}$. As a result, opening a buckling contact produces a small series of uncorrelated displacements. No particular length scale is hence favored over any other. Because of the close relation between localized forces and displacements just mentioned, it is natural for $f_\ell$ to be uniformly distributed.

Before closing this section, we note that the distributions of $f_\ell$ for the FCC and MK packings violate the stability condition related to local excitations given by Eq.~\eqref{eq:gamma_theta_loc}. We comment further on this point in Sec.~\ref{sec:discussion}. For now, we simply note that broader classes of disorder need to be considered when studying the criticality associated with localized contact forces, even though their finite-size effects are unimportant.

\begin{table*}
	\centering
	\renewcommand{\arraystretch}{1.2}
	\begin{tabular}{ V{3} c V{2} c| c| c| c V{2}}
		\hlineB{3}
		Property & $d=3$ Monodisperse & $d=2$ Polydisperse & FCC & MK\\[1mm]
		& UC and OC & OC & OC & UC \\[1mm]
		\hlineB{2.5}
		$p(f_e)$ with $\theta_e=0.42311$ & \cmark & \cmark & \cmark (but small range) & \cmark\\[1mm]
		\hline
		$g(h)$ with $\gamma=0.41629$ & \cmark & \cmark  & {\color{red}\xmark: $\gamma_{FCC} \simeq 0.33$} & \cmark
		\\[1mm]
		\hline
		$p(f_\ell)$ with $\theta_\ell=0.17$ & \cmark & \cmark & {\color{red} \xmark: no power law} & {\color{red}\xmark:  $\theta_{\ell, MK}=0$}\\[1mm]
		\hline
		Eq.~\eqref{eq:scaling-cdf} scaling for forces & \xmark & \xmark & \xmark & \cmark (but small effect)\\[1mm]
		\hline
		Eq.~\eqref{eq:scaling-cdf} scaling for gaps & \cmark & \cmark (Eq.~\eqref{eq:scaling-cdf-log-cor}) & \cmark {\color{red} (using $\gamma_{FCC})$} & \cmark \\[1mm]
		\hlineB{3}
	\end{tabular}
	\caption{Summary of our main results for the various properties and models considered. In the heading we also indicate if the respective jamming point was reached from the under- (UC) or over-compressed (OC) phase. In the first three rows a check-mark (\cmark) denotes that the corresponding theoretical prediction was verified, and a cross (\xmark) that it was not. In the last two rows symbols denote whether the size scaling was verified or not.  Results that contradict MF predictions, or results from previous studies, are highlighted in red.
	}
	\label{tab:results}
\end{table*}

\section{Discussion}\label{sec:discussion}

For clarity, we synthesize our results in Table \ref{tab:results}. The first three rows, which consider the power-law scaling of the pdfs in Eqs.~\eqref{eq:pdf-gaps} and \eqref{eq:pdf-forces}, assess the jamming criticality associated with microstructural variables for different types of systems. Recall that not only were different models considered, but so was the direction of approach to the jamming point. The systematic corroboration of the nontrivial distributions of forces and gaps for fully disordered systems at jamming completely supports the description derived from the exact MF theory. Systems with an underlying FCC symmetry, however, exhibit marked discrepancies. Our result thus validate earlier reports that crystalline structures fall outside the jamming universality~\cite{gardner-crystals,tsekenis_jamming_2020} , even though some of its critical features are conserved~\cite{ikeda_jamming_crystals}.

Our main finding is the contrasted system-size dependence of the distribution of gaps and contact forces, as summarized in the last two rows of Table~\ref{tab:results}. Size effects in $p(f_e)$ are practically nonexistent for all models, dimensionality, and interaction type, while $g(h)$ exhibits clear and systematic signatures of finite-$N$ deviations from the expected power-law scaling. Logarithmic corrections to $g(h)$ are further observed in two-dimensional systems.
We emphasize that testing for such size scalings not only rigorously assesses the critical scaling and its exponents~\cite{amit2005field,MC-book}, but also provides key insight into the length scale of their correlations.
Hence, we conclude that the MF exponents for all gap distributions and the $f_e$ one in the MK model are correct. Yet--leaving aside for the moment the MK results--a second and more informative conclusion is that the distribution $p(f_e)$ reaches its thermodynamic limit behavior at smaller values of $N$ than $g(h)$. Two different correlation lengths, $\xi_{f_e}$ and $\xi_h$, therefore characterize the relevant length scales of correlations of contact forces and gaps, respectively. This finding is rather unexpected, because the critical behavior of both quantities is controlled by the onset of isostaticity at the jamming transition. Moreover, theoretical approaches \cite{puz-book,exact_mft_3,jamming_sat_unsat} suggest that forces and gaps can be studied from a unified viewpoint (essentially by considering forces as the zero limit of negative gaps), and thus they should share a common correlation length, $\xi$. Naturally, in the thermodynamic limit $\xi$ should diverge at the jamming transition, thus signaling system-wide correlations between microscopic variables. Our results for finite-size systems, by contrast, suggest that correlations in gaps and forces have different length scales, namely $\xi_h \gtrsim N^{1/d} \gg \xi_{f_e}$. The fact that no known relation for $\xi$ has been put forward (nor for $\xi_h$ or $\xi_{f_e}$ for that matter) partly obfuscates further analysis. A simple resolution could be to assume that both $\xi_h$ and $\xi_{f_e}$ are proportional to $\xi$, but with a prefactor that is much larger for the former than for the latter.  
Considering that forces and gaps are usually treated on an equal footing from the perspective of the SAT-UNSAT transition in the perceptron~\cite{simplest_jamming,franz_critical_perceptron}, constraint satisfaction problems~\cite{jamming_sat_unsat,MF_non_spheres}, and neural networks~\cite{jamming_multilayer_supervised} as well as from the point of view of marginal stability in amorphous solids~\cite{wyart_marginal_stability,wyart_muller_review_marginal_2015}, the disparity in their correlation lengths is nevertheless surprising.

The MK results also fit into this description if we consider that their very high densities and connectivity reduce the effective system size, as discussed in Sec.~\ref{sec:other-systems}.
Observing the scaling of Eq.~\eqref{eq:scaling-cdf} for the cdf of $f_e$ is thus a manifestation of the smaller effective volume (for a similar $N$), which confirms that finite-size corrections for $p(f_e)$ are present at jamming, but disappear for relatively small system sizes. The significantly more pronounced $N$ dependence of the distributions of $h$ [Figs.~\ref{fig:forces-other-systems}(c) and \ref{fig:gaps-MK}(b)] thus supports our finding that $\xi_h \gg \xi_{f_e}$.

Interestingly, our results further suggest that the marginal stability bounds for the exponents, as expressed in Eqs.~\eqref{eq:marginality-bounds}, should be modified when different types of disorder are present. For instance, our findings along with other works~\cite{gardner-crystals,tsekenis_jamming_2020} evince that these inequalities are prone to be violated when crystalline lattices are used to generate the jammed packings.  The inherent geometry of jammed configurations therefore plays a significant role in formulating general stability criteria.  Because the bounds in Eq.~\eqref{eq:marginality-bounds} were derived~\cite{wyart_marginal_stability,lerner_low-energy_excitations,wyart_muller_review_marginal_2015} assuming, implicitly, that particles positions are uncorrelated, it should not be overly surprising that $\gamma_{FCC}$ violates both relations. It nevertheless suggests that, despite being likewise composed of frictionless spheres, near-crystals are not part of the same universality class.

The linear growth of cdf$(f_\ell)$ in the MK model is also at odds with the stability condition of Eq.~\eqref{eq:gamma_theta_loc}. This finding is more surprising because there is no long-range order in this type of system. At the end of Sec.~\ref{sec:loc-forces} we used the peculiar geometry of these packings to suggest a physical explanation for the uniform distribution of $f_\ell$, but this reasoning does not explain why the stability condition between $\gamma$ and $\theta_\ell$ is apparently violated. Given the drastic difference in the inherent structures of the FCC and MK packings, they highlight the need for more studies to better understand the role played by disorder in determining how the response to external perturbations is related to spatial correlations between particles in jammed systems.

The most persistent observation was that all cumulative distributions of both gaps and extended forces behave in a seemingly linear fashion at very small arguments, in agreement with the MF predictions, $p_0$ and $g_0$ in Eqs.~\eqref{eq:pdf-gaps-v2} and \eqref{eq:pdf-f-ext-v2}, respectively. Such a cutoff of the main power-law scaling is due to the extra contact of isostatic configurations and its effect in the scaling function can be captured using the same scaling transformation we performed for the main power-law scaling (see Sec.~\ref{sec:expected scalings}, especially Eq.~\eqref{eq:general scaling cdf}). It has been previously reported for the gaps distributions of disks packings~\cite{ikeda_infinitesimal_2020}, but we are not aware of analogous findings in any other model or for the $f_e$ distributions. As discussed at the end of Sec.~\ref{sec:other-systems}, our results suggest that scalings caused by the additional contact with respect to isostaticity are more robust against changes in the type of disorder and have a similar characteristic scale in both types of microstructural variables. However, because of undersampling of the left tails of these distributions, a more stringent analysis would need to be carried out to verify that $p_0(x) \sim g_0(x) \sim 1$ when $x\ll1$.
A previous work on the perceptron~\cite{perceptron_size_scaling} also reported a similar transition to a uniform distribution of contact forces that depended on the type of algorithm used to reach the jamming point, but given that we have used two different algorithms to produce our packings, it is unlikely that both could produce the same systematic effect. This question is particularly interesting because it would directly affect the robustness of jamming universality, albeit only for the very smallest forces and gaps. Yet, given that the left tails of $g(h)$ and $p(f_e)$ determine the smallest gaps and contact forces, accurately describing their true distribution is  key to assessing the stability of jammed packings away from the thermodynamic limit. We nevertheless leave this and other related issues as topics for future consideration.\\[5mm]

Data relevant to this work have been archived and can be accessed from the Duke Digital Repository \footnote{Duke Digital Repository: \url{ https://doi.org/10.7924/r4833vm1m}}.

\acknowledgments
\noindent We want to thank Franceso Zamponi for insightful comments and suggestions to our work. RDHR thanks Georgios Tsekenis for very useful discussions during the initial stage of this work and Beatriz Seoane for helpful suggestions regarding the molecular dynamics simulations of the MK model. This work was supported by the Simons Foundation grant (\# 454937, PC; \# 454939, EC; \# 454949 GP) as well as by the European Research Council under the European Unions Horizon 2020 research and innovation program (grant No. 694925, G.P.). H.I. was supported by JSPS KAKENHI No. 20J00289.

\bibliography{refs-size-effects-jamming.bib}

\begin{thebibliography}{80}%
\makeatletter
\providecommand \@ifxundefined [1]{%
 \@ifx{#1\undefined}
}%
\providecommand \@ifnum [1]{%
 \ifnum #1\expandafter \@firstoftwo
 \else \expandafter \@secondoftwo
 \fi
}%
\providecommand \@ifx [1]{%
 \ifx #1\expandafter \@firstoftwo
 \else \expandafter \@secondoftwo
 \fi
}%
\providecommand \natexlab [1]{#1}%
\providecommand \enquote  [1]{``#1''}%
\providecommand \bibnamefont  [1]{#1}%
\providecommand \bibfnamefont [1]{#1}%
\providecommand \citenamefont [1]{#1}%
\providecommand \href@noop [0]{\@secondoftwo}%
\providecommand \href [0]{\begingroup \@sanitize@url \@href}%
\providecommand \@href[1]{\@@startlink{#1}\@@href}%
\providecommand \@@href[1]{\endgroup#1\@@endlink}%
\providecommand \@sanitize@url [0]{\catcode `\\12\catcode `\$12\catcode
  `\&12\catcode `\#12\catcode `\^12\catcode `\_12\catcode `\%12\relax}%
\providecommand \@@startlink[1]{}%
\providecommand \@@endlink[0]{}%
\providecommand \url  [0]{\begingroup\@sanitize@url \@url }%
\providecommand \@url [1]{\endgroup\@href {#1}{\urlprefix }}%
\providecommand \urlprefix  [0]{URL }%
\providecommand \Eprint [0]{\href }%
\providecommand \doibase [0]{http://dx.doi.org/}%
\providecommand \selectlanguage [0]{\@gobble}%
\providecommand \bibinfo  [0]{\@secondoftwo}%
\providecommand \bibfield  [0]{\@secondoftwo}%
\providecommand \translation [1]{[#1]}%
\providecommand \BibitemOpen [0]{}%
\providecommand \bibitemStop [0]{}%
\providecommand \bibitemNoStop [0]{.\EOS\space}%
\providecommand \EOS [0]{\spacefactor3000\relax}%
\providecommand \BibitemShut  [1]{\csname bibitem#1\endcsname}%
\let\auto@bib@innerbib\@empty
\bibitem [{\citenamefont {Parisi}\ \emph {et~al.}(2020)\citenamefont {Parisi},
  \citenamefont {Urbani},\ and\ \citenamefont {Zamponi}}]{puz-book}%
  \BibitemOpen
  \bibfield  {author} {\bibinfo {author} {\bibfnamefont {Giorgio}\ \bibnamefont
  {Parisi}}, \bibinfo {author} {\bibfnamefont {Pierfrancesco}\ \bibnamefont
  {Urbani}}, \ and\ \bibinfo {author} {\bibfnamefont {Francesco}\ \bibnamefont
  {Zamponi}},\ }\href {\doibase 10.1017/9781108120494} {\emph {\bibinfo {title}
  {Theory of {{Simple Glasses}}: {{Exact Solutions}} in {{Infinite
  Dimensions}}}}}\ (\bibinfo  {publisher} {{Cambridge University Press}},\
  \bibinfo {address} {{Cambridge}},\ \bibinfo {year} {2020})\BibitemShut
  {NoStop}%
\bibitem [{\citenamefont {Berthier}\ and\ \citenamefont
  {Biroli}(2011)}]{berthier_biroli-review-2011}%
  \BibitemOpen
  \bibfield  {author} {\bibinfo {author} {\bibfnamefont {Ludovic}\ \bibnamefont
  {Berthier}}\ and\ \bibinfo {author} {\bibfnamefont {Giulio}\ \bibnamefont
  {Biroli}},\ }\bibfield  {title} {\enquote {\bibinfo {title} {Theoretical
  perspective on the glass transition and amorphous materials},}\ }\href
  {\doibase 10.1103/RevModPhys.83.587} {\bibfield  {journal} {\bibinfo
  {journal} {Reviews of Modern Physics}\ }\textbf {\bibinfo {volume} {83}},\
  \bibinfo {pages} {587--645} (\bibinfo {year} {2011})}\BibitemShut {NoStop}%
\bibitem [{\citenamefont {Liu}\ and\ \citenamefont
  {Nagel}(2010)}]{liu_nagel_review_2010}%
  \BibitemOpen
  \bibfield  {author} {\bibinfo {author} {\bibfnamefont {Andrea~J.}\
  \bibnamefont {Liu}}\ and\ \bibinfo {author} {\bibfnamefont {Sidney~R.}\
  \bibnamefont {Nagel}},\ }\bibfield  {title} {\enquote {\bibinfo {title} {The
  {Jamming} {Transition} and the {Marginally} {Jammed} {Solid}},}\ }\href
  {\doibase 10.1146/annurev-conmatphys-070909-104045} {\bibfield  {journal}
  {\bibinfo  {journal} {Annual Review of Condensed Matter Physics}\ }\textbf
  {\bibinfo {volume} {1}},\ \bibinfo {pages} {347--369} (\bibinfo {year}
  {2010})}\BibitemShut {NoStop}%
\bibitem [{\citenamefont {Parisi}\ and\ \citenamefont
  {Zamponi}(2010)}]{mft-review_2010}%
  \BibitemOpen
  \bibfield  {author} {\bibinfo {author} {\bibfnamefont {Giorgio}\ \bibnamefont
  {Parisi}}\ and\ \bibinfo {author} {\bibfnamefont {Francesco}\ \bibnamefont
  {Zamponi}},\ }\bibfield  {title} {\enquote {\bibinfo {title} {Mean-field
  theory of hard sphere glasses and jamming},}\ }\href {\doibase
  10.1103/RevModPhys.82.789} {\bibfield  {journal} {\bibinfo  {journal}
  {Reviews of Modern Physics}\ }\textbf {\bibinfo {volume} {82}},\ \bibinfo
  {pages} {789--845} (\bibinfo {year} {2010})}\BibitemShut {NoStop}%
\bibitem [{\citenamefont {Torquato}\ and\ \citenamefont
  {Stillinger}(2010)}]{torquato_review_2010}%
  \BibitemOpen
  \bibfield  {author} {\bibinfo {author} {\bibfnamefont {S.}~\bibnamefont
  {Torquato}}\ and\ \bibinfo {author} {\bibfnamefont {F.~H.}\ \bibnamefont
  {Stillinger}},\ }\bibfield  {title} {\enquote {\bibinfo {title} {Jammed
  hard-particle packings: {From} {Kepler} to {Bernal} and beyond},}\ }\href
  {\doibase 10.1103/RevModPhys.82.2633} {\bibfield  {journal} {\bibinfo
  {journal} {Reviews of Modern Physics}\ }\textbf {\bibinfo {volume} {82}},\
  \bibinfo {pages} {2633--2672} (\bibinfo {year} {2010})}\BibitemShut {NoStop}%
\bibitem [{\citenamefont {Charbonneau}\ \emph {et~al.}(2017)\citenamefont
  {Charbonneau}, \citenamefont {Kurchan}, \citenamefont {Parisi}, \citenamefont
  {Urbani},\ and\ \citenamefont {Zamponi}}]{exact_mft_review}%
  \BibitemOpen
  \bibfield  {author} {\bibinfo {author} {\bibfnamefont {Patrick}\ \bibnamefont
  {Charbonneau}}, \bibinfo {author} {\bibfnamefont {Jorge}\ \bibnamefont
  {Kurchan}}, \bibinfo {author} {\bibfnamefont {Giorgio}\ \bibnamefont
  {Parisi}}, \bibinfo {author} {\bibfnamefont {Pierfrancesco}\ \bibnamefont
  {Urbani}}, \ and\ \bibinfo {author} {\bibfnamefont {Francesco}\ \bibnamefont
  {Zamponi}},\ }\bibfield  {title} {\enquote {\bibinfo {title} {Glass and
  {Jamming} {Transitions}: {From} {Exact} {Results} to {Finite}-{Dimensional}
  {Descriptions}},}\ }\href {\doibase 10.1146/annurev-conmatphys-031016-025334}
  {\bibfield  {journal} {\bibinfo  {journal} {Annual Review of Condensed Matter
  Physics}\ }\textbf {\bibinfo {volume} {8}},\ \bibinfo {pages} {265--288}
  (\bibinfo {year} {2017})}\BibitemShut {NoStop}%
\bibitem [{\citenamefont {van Hecke}(2010)}]{jamming_van_hecke}%
  \BibitemOpen
  \bibfield  {author} {\bibinfo {author} {\bibfnamefont {Martin}\ \bibnamefont
  {van Hecke}},\ }\bibfield  {title} {\enquote {\bibinfo {title} {Jamming of
  soft particles: geometry, mechanics, scaling and isostaticity},}\ }\href
  {\doibase 10.1088/0953-8984/22/3/033101} {\bibfield  {journal} {\bibinfo
  {journal} {Journal of Physics: Condensed Matter}\ }\textbf {\bibinfo {volume}
  {22}},\ \bibinfo {pages} {033101} (\bibinfo {year} {2010})}\BibitemShut
  {NoStop}%
\bibitem [{\citenamefont {Baule}\ \emph {et~al.}(2018)\citenamefont {Baule},
  \citenamefont {Morone}, \citenamefont {Herrmann},\ and\ \citenamefont
  {Makse}}]{review_edwards_statmech-jamming}%
  \BibitemOpen
  \bibfield  {author} {\bibinfo {author} {\bibfnamefont {Adrian}\ \bibnamefont
  {Baule}}, \bibinfo {author} {\bibfnamefont {Flaviano}\ \bibnamefont
  {Morone}}, \bibinfo {author} {\bibfnamefont {Hans~J.}\ \bibnamefont
  {Herrmann}}, \ and\ \bibinfo {author} {\bibfnamefont {Hernán~A.}\
  \bibnamefont {Makse}},\ }\bibfield  {title} {\enquote {\bibinfo {title}
  {Edwards statistical mechanics for jammed granular matter},}\ }\href
  {\doibase 10.1103/RevModPhys.90.015006} {\bibfield  {journal} {\bibinfo
  {journal} {Reviews of Modern Physics}\ }\textbf {\bibinfo {volume} {90}},\
  \bibinfo {pages} {015006} (\bibinfo {year} {2018})}\BibitemShut {NoStop}%
\bibitem [{\citenamefont {Liu}\ and\ \citenamefont
  {Nagel}(1998)}]{liu_nagel_1998}%
  \BibitemOpen
  \bibfield  {author} {\bibinfo {author} {\bibfnamefont {Andrea~J.}\
  \bibnamefont {Liu}}\ and\ \bibinfo {author} {\bibfnamefont {Sidney~R.}\
  \bibnamefont {Nagel}},\ }\bibfield  {title} {\enquote {\bibinfo {title}
  {Nonlinear dynamics: {Jamming} is not just cool any more},}\ }\href {\doibase
  10.1038/23819} {\bibfield  {journal} {\bibinfo  {journal} {Nature}\ }\textbf
  {\bibinfo {volume} {396}},\ \bibinfo {pages} {21--22} (\bibinfo {year}
  {1998})}\BibitemShut {NoStop}%
\bibitem [{\citenamefont {O’Hern}\ \emph {et~al.}(2003)\citenamefont
  {O’Hern}, \citenamefont {Silbert}, \citenamefont {Liu},\ and\ \citenamefont
  {Nagel}}]{ohern_liu_2003}%
  \BibitemOpen
  \bibfield  {author} {\bibinfo {author} {\bibfnamefont {Corey~S.}\
  \bibnamefont {O’Hern}}, \bibinfo {author} {\bibfnamefont {Leonardo~E.}\
  \bibnamefont {Silbert}}, \bibinfo {author} {\bibfnamefont {Andrea~J.}\
  \bibnamefont {Liu}}, \ and\ \bibinfo {author} {\bibfnamefont {Sidney~R.}\
  \bibnamefont {Nagel}},\ }\bibfield  {title} {\enquote {\bibinfo {title}
  {Jamming at zero temperature and zero applied stress: {The} epitome of
  disorder},}\ }\href {\doibase 10.1103/PhysRevE.68.011306} {\bibfield
  {journal} {\bibinfo  {journal} {Physical Review E}\ }\textbf {\bibinfo
  {volume} {68}},\ \bibinfo {pages} {011306} (\bibinfo {year}
  {2003})}\BibitemShut {NoStop}%
\bibitem [{\citenamefont {Charbonneau}\ \emph {et~al.}(2012)\citenamefont
  {Charbonneau}, \citenamefont {Corwin}, \citenamefont {Parisi},\ and\
  \citenamefont {Zamponi}}]{universal_microstructure}%
  \BibitemOpen
  \bibfield  {author} {\bibinfo {author} {\bibfnamefont {Patrick}\ \bibnamefont
  {Charbonneau}}, \bibinfo {author} {\bibfnamefont {Eric~I.}\ \bibnamefont
  {Corwin}}, \bibinfo {author} {\bibfnamefont {Giorgio}\ \bibnamefont
  {Parisi}}, \ and\ \bibinfo {author} {\bibfnamefont {Francesco}\ \bibnamefont
  {Zamponi}},\ }\bibfield  {title} {\enquote {\bibinfo {title} {Universal
  {Microstructure} and {Mechanical} {Stability} of {Jammed} {Packings}},}\
  }\href {\doibase 10.1103/PhysRevLett.109.205501} {\bibfield  {journal}
  {\bibinfo  {journal} {Physical Review Letters}\ }\textbf {\bibinfo {volume}
  {109}},\ \bibinfo {pages} {205501} (\bibinfo {year} {2012})}\BibitemShut
  {NoStop}%
\bibitem [{\citenamefont {Kurchan}\ \emph {et~al.}(2012)\citenamefont
  {Kurchan}, \citenamefont {Parisi},\ and\ \citenamefont
  {Zamponi}}]{exact_mft_1}%
  \BibitemOpen
  \bibfield  {author} {\bibinfo {author} {\bibfnamefont {Jorge}\ \bibnamefont
  {Kurchan}}, \bibinfo {author} {\bibfnamefont {Giorgio}\ \bibnamefont
  {Parisi}}, \ and\ \bibinfo {author} {\bibfnamefont {Francesco}\ \bibnamefont
  {Zamponi}},\ }\bibfield  {title} {\enquote {\bibinfo {title} {Exact theory of
  dense amorphous hard spheres in high dimension {I}. {The} free energy},}\
  }\href {\doibase 10.1088/1742-5468/2012/10/P10012} {\bibfield  {journal}
  {\bibinfo  {journal} {Journal of Statistical Mechanics: Theory and
  Experiment}\ }\textbf {\bibinfo {volume} {2012}},\ \bibinfo {pages} {P10012}
  (\bibinfo {year} {2012})}\BibitemShut {NoStop}%
\bibitem [{\citenamefont {Kurchan}\ \emph {et~al.}(2013)\citenamefont
  {Kurchan}, \citenamefont {Parisi}, \citenamefont {Urbani},\ and\
  \citenamefont {Zamponi}}]{exact_mft_2}%
  \BibitemOpen
  \bibfield  {author} {\bibinfo {author} {\bibfnamefont {Jorge}\ \bibnamefont
  {Kurchan}}, \bibinfo {author} {\bibfnamefont {Giorgio}\ \bibnamefont
  {Parisi}}, \bibinfo {author} {\bibfnamefont {Pierfrancesco}\ \bibnamefont
  {Urbani}}, \ and\ \bibinfo {author} {\bibfnamefont {Francesco}\ \bibnamefont
  {Zamponi}},\ }\bibfield  {title} {\enquote {\bibinfo {title} {Exact {Theory}
  of {Dense} {Amorphous} {Hard} {Spheres} in {High} {Dimension}. {II}. {The}
  {High} {Density} {Regime} and the {Gardner} {Transition}},}\ }\href {\doibase
  10.1021/jp402235d} {\bibfield  {journal} {\bibinfo  {journal} {The Journal of
  Physical Chemistry B}\ }\textbf {\bibinfo {volume} {117}},\ \bibinfo {pages}
  {12979--12994} (\bibinfo {year} {2013})}\BibitemShut {NoStop}%
\bibitem [{\citenamefont {Charbonneau}\ \emph
  {et~al.}(2014{\natexlab{a}})\citenamefont {Charbonneau}, \citenamefont
  {Kurchan}, \citenamefont {Parisi}, \citenamefont {Urbani},\ and\
  \citenamefont {Zamponi}}]{exact_mft_3}%
  \BibitemOpen
  \bibfield  {author} {\bibinfo {author} {\bibfnamefont {Patrick}\ \bibnamefont
  {Charbonneau}}, \bibinfo {author} {\bibfnamefont {Jorge}\ \bibnamefont
  {Kurchan}}, \bibinfo {author} {\bibfnamefont {Giorgio}\ \bibnamefont
  {Parisi}}, \bibinfo {author} {\bibfnamefont {Pierfrancesco}\ \bibnamefont
  {Urbani}}, \ and\ \bibinfo {author} {\bibfnamefont {Francesco}\ \bibnamefont
  {Zamponi}},\ }\bibfield  {title} {\enquote {\bibinfo {title} {Exact theory of
  dense amorphous hard spheres in high dimension. {III}. {The} full replica
  symmetry breaking solution},}\ }\href {\doibase
  10.1088/1742-5468/2014/10/P10009} {\bibfield  {journal} {\bibinfo  {journal}
  {Journal of Statistical Mechanics: Theory and Experiment}\ }\textbf {\bibinfo
  {volume} {2014}},\ \bibinfo {pages} {P10009} (\bibinfo {year}
  {2014}{\natexlab{a}})}\BibitemShut {NoStop}%
\bibitem [{\citenamefont {Charbonneau}\ \emph
  {et~al.}(2014{\natexlab{b}})\citenamefont {Charbonneau}, \citenamefont
  {Kurchan}, \citenamefont {Parisi}, \citenamefont {Urbani},\ and\
  \citenamefont {Zamponi}}]{fractal_fel}%
  \BibitemOpen
  \bibfield  {author} {\bibinfo {author} {\bibfnamefont {Patrick}\ \bibnamefont
  {Charbonneau}}, \bibinfo {author} {\bibfnamefont {Jorge}\ \bibnamefont
  {Kurchan}}, \bibinfo {author} {\bibfnamefont {Giorgio}\ \bibnamefont
  {Parisi}}, \bibinfo {author} {\bibfnamefont {Pierfrancesco}\ \bibnamefont
  {Urbani}}, \ and\ \bibinfo {author} {\bibfnamefont {Francesco}\ \bibnamefont
  {Zamponi}},\ }\bibfield  {title} {\enquote {\bibinfo {title} {Fractal free
  energy landscapes in structural glasses},}\ }\href {\doibase
  10.1038/ncomms4725} {\bibfield  {journal} {\bibinfo  {journal} {Nature
  Communications}\ }\textbf {\bibinfo {volume} {5}},\ \bibinfo {pages} {3725}
  (\bibinfo {year} {2014}{\natexlab{b}})}\BibitemShut {NoStop}%
\bibitem [{\citenamefont {Charbonneau}\ \emph
  {et~al.}(2015{\natexlab{a}})\citenamefont {Charbonneau}, \citenamefont
  {Corwin}, \citenamefont {Parisi},\ and\ \citenamefont
  {Zamponi}}]{jamming_criticality_rem_bucks}%
  \BibitemOpen
  \bibfield  {author} {\bibinfo {author} {\bibfnamefont {Patrick}\ \bibnamefont
  {Charbonneau}}, \bibinfo {author} {\bibfnamefont {Eric~I.}\ \bibnamefont
  {Corwin}}, \bibinfo {author} {\bibfnamefont {Giorgio}\ \bibnamefont
  {Parisi}}, \ and\ \bibinfo {author} {\bibfnamefont {Francesco}\ \bibnamefont
  {Zamponi}},\ }\bibfield  {title} {\enquote {\bibinfo {title} {Jamming
  {Criticality} {Revealed} by {Removing} {Localized} {Buckling}
  {Excitations}},}\ }\href {\doibase 10.1103/PhysRevLett.114.125504} {\bibfield
   {journal} {\bibinfo  {journal} {Physical Review Letters}\ }\textbf {\bibinfo
  {volume} {114}},\ \bibinfo {pages} {125504} (\bibinfo {year}
  {2015}{\natexlab{a}})}\BibitemShut {NoStop}%
\bibitem [{\citenamefont {Lerner}\ \emph {et~al.}(2013)\citenamefont {Lerner},
  \citenamefont {Düring},\ and\ \citenamefont
  {Wyart}}]{lerner_low-energy_excitations}%
  \BibitemOpen
  \bibfield  {author} {\bibinfo {author} {\bibfnamefont {Edan}\ \bibnamefont
  {Lerner}}, \bibinfo {author} {\bibfnamefont {Gustavo}\ \bibnamefont
  {Düring}}, \ and\ \bibinfo {author} {\bibfnamefont {Matthieu}\ \bibnamefont
  {Wyart}},\ }\bibfield  {title} {\enquote {\bibinfo {title} {Low-energy
  non-linear excitations in sphere packings},}\ }\href {\doibase
  10.1039/C3SM50515D} {\bibfield  {journal} {\bibinfo  {journal} {Soft Matter}\
  }\textbf {\bibinfo {volume} {9}},\ \bibinfo {pages} {8252--8263} (\bibinfo
  {year} {2013})}\BibitemShut {NoStop}%
\bibitem [{\citenamefont {Ikeda}(2020{\natexlab{a}})}]{ikeda2020}%
  \BibitemOpen
  \bibfield  {author} {\bibinfo {author} {\bibfnamefont {Harukuni}\
  \bibnamefont {Ikeda}},\ }\bibfield  {title} {\enquote {\bibinfo {title}
  {Jamming below upper critical dimension},}\ }\href {\doibase
  10.1103/PhysRevLett.125.038001} {\bibfield  {journal} {\bibinfo  {journal}
  {Physical Review Letters}\ }\textbf {\bibinfo {volume} {125}},\ \bibinfo
  {pages} {038001} (\bibinfo {year} {2020}{\natexlab{a}})}\BibitemShut
  {NoStop}%
\bibitem [{\citenamefont {Zhang}\ \emph {et~al.}(2020)\citenamefont {Zhang},
  \citenamefont {Godfrey},\ and\ \citenamefont
  {Moore}}]{moore_marginally_2020}%
  \BibitemOpen
  \bibfield  {author} {\bibinfo {author} {\bibfnamefont {Yuxiao}\ \bibnamefont
  {Zhang}}, \bibinfo {author} {\bibfnamefont {M.~J.}\ \bibnamefont {Godfrey}},
  \ and\ \bibinfo {author} {\bibfnamefont {M.~A.}\ \bibnamefont {Moore}},\
  }\bibfield  {title} {\enquote {\bibinfo {title} {Marginally jammed states of
  hard disks in a one-dimensional channel},}\ }\href
  {http://arxiv.org/abs/2005.02901} {\bibfield  {journal} {\bibinfo  {journal}
  {arXiv:2005.02901 [cond-mat]}\ } (\bibinfo {year} {2020})}\BibitemShut
  {NoStop}%
\bibitem [{\citenamefont {Ikeda}\ \emph {et~al.}(2013)\citenamefont {Ikeda},
  \citenamefont {Berthier},\ and\ \citenamefont
  {Biroli}}]{jamming-dynamic-criticality}%
  \BibitemOpen
  \bibfield  {author} {\bibinfo {author} {\bibfnamefont {Atsushi}\ \bibnamefont
  {Ikeda}}, \bibinfo {author} {\bibfnamefont {Ludovic}\ \bibnamefont
  {Berthier}}, \ and\ \bibinfo {author} {\bibfnamefont {Giulio}\ \bibnamefont
  {Biroli}},\ }\bibfield  {title} {\enquote {\bibinfo {title} {Dynamic
  criticality at the jamming transition},}\ }\href {\doibase 10.1063/1.4769251}
  {\bibfield  {journal} {\bibinfo  {journal} {The Journal of Chemical Physics}\
  }\textbf {\bibinfo {volume} {138}},\ \bibinfo {pages} {12A507} (\bibinfo
  {year} {2013})}\BibitemShut {NoStop}%
\bibitem [{\citenamefont {Hopkins}\ \emph {et~al.}(2013)\citenamefont
  {Hopkins}, \citenamefont {Stillinger},\ and\ \citenamefont
  {Torquato}}]{hopkins_disordered_2013}%
  \BibitemOpen
  \bibfield  {author} {\bibinfo {author} {\bibfnamefont {Adam~B.}\ \bibnamefont
  {Hopkins}}, \bibinfo {author} {\bibfnamefont {Frank~H.}\ \bibnamefont
  {Stillinger}}, \ and\ \bibinfo {author} {\bibfnamefont {Salvatore}\
  \bibnamefont {Torquato}},\ }\bibfield  {title} {\enquote {\bibinfo {title}
  {Disordered strictly jammed binary sphere packings attain an anomalously
  large range of densities},}\ }\href {\doibase 10.1103/PhysRevE.88.022205}
  {\bibfield  {journal} {\bibinfo  {journal} {Physical Review E}\ }\textbf
  {\bibinfo {volume} {88}},\ \bibinfo {pages} {022205} (\bibinfo {year}
  {2013})}\BibitemShut {NoStop}%
\bibitem [{\citenamefont {Skoge}\ \emph {et~al.}(2006)\citenamefont {Skoge},
  \citenamefont {Donev}, \citenamefont {Stillinger},\ and\ \citenamefont
  {Torquato}}]{md_ls}%
  \BibitemOpen
  \bibfield  {author} {\bibinfo {author} {\bibfnamefont {Monica}\ \bibnamefont
  {Skoge}}, \bibinfo {author} {\bibfnamefont {Aleksandar}\ \bibnamefont
  {Donev}}, \bibinfo {author} {\bibfnamefont {Frank~H.}\ \bibnamefont
  {Stillinger}}, \ and\ \bibinfo {author} {\bibfnamefont {Salvatore}\
  \bibnamefont {Torquato}},\ }\bibfield  {title} {\enquote {\bibinfo {title}
  {Packing hyperspheres in high-dimensional {Euclidean} spaces},}\ }\href
  {\doibase 10.1103/PhysRevE.74.041127} {\bibfield  {journal} {\bibinfo
  {journal} {Physical Review E}\ }\textbf {\bibinfo {volume} {74}},\ \bibinfo
  {pages} {041127} (\bibinfo {year} {2006})}\BibitemShut {NoStop}%
\bibitem [{\citenamefont {Berthier}\ \emph {et~al.}(2016)\citenamefont
  {Berthier}, \citenamefont {Charbonneau}, \citenamefont {Jin}, \citenamefont
  {Parisi}, \citenamefont {Seoane},\ and\ \citenamefont
  {Zamponi}}]{gardner_pnas_2016}%
  \BibitemOpen
  \bibfield  {author} {\bibinfo {author} {\bibfnamefont {Ludovic}\ \bibnamefont
  {Berthier}}, \bibinfo {author} {\bibfnamefont {Patrick}\ \bibnamefont
  {Charbonneau}}, \bibinfo {author} {\bibfnamefont {Yuliang}\ \bibnamefont
  {Jin}}, \bibinfo {author} {\bibfnamefont {Giorgio}\ \bibnamefont {Parisi}},
  \bibinfo {author} {\bibfnamefont {Beatriz}\ \bibnamefont {Seoane}}, \ and\
  \bibinfo {author} {\bibfnamefont {Francesco}\ \bibnamefont {Zamponi}},\
  }\bibfield  {title} {\enquote {\bibinfo {title} {Growing timescales and
  lengthscales characterizing vibrations of amorphous solids},}\ }\href
  {\doibase 10.1073/pnas.1607730113} {\bibfield  {journal} {\bibinfo  {journal}
  {Proceedings of the National Academy of Sciences}\ }\textbf {\bibinfo
  {volume} {113}},\ \bibinfo {pages} {8397--8401} (\bibinfo {year}
  {2016})}\BibitemShut {NoStop}%
\bibitem [{\citenamefont {Torquato}\ and\ \citenamefont
  {Jiao}(2010)}]{torquatoRobustAlgorithmGenerate2010}%
  \BibitemOpen
  \bibfield  {author} {\bibinfo {author} {\bibfnamefont {S.}~\bibnamefont
  {Torquato}}\ and\ \bibinfo {author} {\bibfnamefont {Y.}~\bibnamefont
  {Jiao}},\ }\bibfield  {title} {\enquote {\bibinfo {title} {Robust algorithm
  to generate a diverse class of dense disordered and ordered sphere packings
  via linear programming},}\ }\href {\doibase 10.1103/PhysRevE.82.061302}
  {\bibfield  {journal} {\bibinfo  {journal} {Physical Review E}\ }\textbf
  {\bibinfo {volume} {82}},\ \bibinfo {pages} {061302} (\bibinfo {year}
  {2010})}\BibitemShut {NoStop}%
\bibitem [{\citenamefont {Jiao}\ \emph {et~al.}(2011)\citenamefont {Jiao},
  \citenamefont {Stillinger},\ and\ \citenamefont
  {Torquato}}]{jiaoNonuniversalityDensityDisorder2011}%
  \BibitemOpen
  \bibfield  {author} {\bibinfo {author} {\bibfnamefont {Yang}\ \bibnamefont
  {Jiao}}, \bibinfo {author} {\bibfnamefont {Frank~H.}\ \bibnamefont
  {Stillinger}}, \ and\ \bibinfo {author} {\bibfnamefont {Salvatore}\
  \bibnamefont {Torquato}},\ }\bibfield  {title} {\enquote {\bibinfo {title}
  {Nonuniversality of density and disorder in jammed sphere packings},}\ }\href
  {\doibase 10.1063/1.3524489} {\bibfield  {journal} {\bibinfo  {journal}
  {Journal of Applied Physics}\ }\textbf {\bibinfo {volume} {109}},\ \bibinfo
  {pages} {013508} (\bibinfo {year} {2011})}\BibitemShut {NoStop}%
\bibitem [{\citenamefont {DeGiuli}\ \emph {et~al.}(2015)\citenamefont
  {DeGiuli}, \citenamefont {Lerner},\ and\ \citenamefont
  {Wyart}}]{jamming-transition-temperature}%
  \BibitemOpen
  \bibfield  {author} {\bibinfo {author} {\bibfnamefont {E.}~\bibnamefont
  {DeGiuli}}, \bibinfo {author} {\bibfnamefont {E.}~\bibnamefont {Lerner}}, \
  and\ \bibinfo {author} {\bibfnamefont {M.}~\bibnamefont {Wyart}},\ }\bibfield
   {title} {\enquote {\bibinfo {title} {Theory of the jamming transition at
  finite temperature},}\ }\href {\doibase 10.1063/1.4918737} {\bibfield
  {journal} {\bibinfo  {journal} {The Journal of Chemical Physics}\ }\textbf
  {\bibinfo {volume} {142}},\ \bibinfo {pages} {164503} (\bibinfo {year}
  {2015})}\BibitemShut {NoStop}%
\bibitem [{\citenamefont {Moukarzel}(1998)}]{moukarzel_isostatic_1998}%
  \BibitemOpen
  \bibfield  {author} {\bibinfo {author} {\bibfnamefont {Cristian~F.}\
  \bibnamefont {Moukarzel}},\ }\bibfield  {title} {\enquote {\bibinfo {title}
  {Isostatic {Phase} {Transition} and {Instability} in {Stiff} {Granular}
  {Materials}},}\ }\href {\doibase 10.1103/PhysRevLett.81.1634} {\bibfield
  {journal} {\bibinfo  {journal} {Physical Review Letters}\ }\textbf {\bibinfo
  {volume} {81}},\ \bibinfo {pages} {1634--1637} (\bibinfo {year}
  {1998})}\BibitemShut {NoStop}%
\bibitem [{\citenamefont {Goodrich}\ \emph {et~al.}(2012)\citenamefont
  {Goodrich}, \citenamefont {Liu},\ and\ \citenamefont
  {Nagel}}]{liu_size_scaling}%
  \BibitemOpen
  \bibfield  {author} {\bibinfo {author} {\bibfnamefont {Carl~P.}\ \bibnamefont
  {Goodrich}}, \bibinfo {author} {\bibfnamefont {Andrea~J.}\ \bibnamefont
  {Liu}}, \ and\ \bibinfo {author} {\bibfnamefont {Sidney~R.}\ \bibnamefont
  {Nagel}},\ }\bibfield  {title} {\enquote {\bibinfo {title} {Finite-{Size}
  {Scaling} at the {Jamming} {Transition}},}\ }\href {\doibase
  10.1103/PhysRevLett.109.095704} {\bibfield  {journal} {\bibinfo  {journal}
  {Physical Review Letters}\ }\textbf {\bibinfo {volume} {109}},\ \bibinfo
  {pages} {095704} (\bibinfo {year} {2012})}\BibitemShut {NoStop}%
\bibitem [{\citenamefont {Goodrich}\ \emph {et~al.}(2014)\citenamefont
  {Goodrich}, \citenamefont {Dagois-Bohy}, \citenamefont {Tighe}, \citenamefont
  {van Hecke}, \citenamefont {Liu},\ and\ \citenamefont
  {Nagel}}]{liu_jamming_finite_systems}%
  \BibitemOpen
  \bibfield  {author} {\bibinfo {author} {\bibfnamefont {Carl~P.}\ \bibnamefont
  {Goodrich}}, \bibinfo {author} {\bibfnamefont {Simon}\ \bibnamefont
  {Dagois-Bohy}}, \bibinfo {author} {\bibfnamefont {Brian~P.}\ \bibnamefont
  {Tighe}}, \bibinfo {author} {\bibfnamefont {Martin}\ \bibnamefont {van
  Hecke}}, \bibinfo {author} {\bibfnamefont {Andrea~J.}\ \bibnamefont {Liu}}, \
  and\ \bibinfo {author} {\bibfnamefont {Sidney~R.}\ \bibnamefont {Nagel}},\
  }\bibfield  {title} {\enquote {\bibinfo {title} {Jamming in finite systems:
  {Stability}, anisotropy, fluctuations, and scaling},}\ }\href {\doibase
  10.1103/PhysRevE.90.022138} {\bibfield  {journal} {\bibinfo  {journal}
  {Physical Review E}\ }\textbf {\bibinfo {volume} {90}},\ \bibinfo {pages}
  {022138} (\bibinfo {year} {2014})}\BibitemShut {NoStop}%
\bibitem [{\citenamefont {Goodrich}\ \emph {et~al.}(2016)\citenamefont
  {Goodrich}, \citenamefont {Liu},\ and\ \citenamefont
  {Sethna}}]{scaling_ansatz_jamming}%
  \BibitemOpen
  \bibfield  {author} {\bibinfo {author} {\bibfnamefont {Carl~P.}\ \bibnamefont
  {Goodrich}}, \bibinfo {author} {\bibfnamefont {Andrea~J.}\ \bibnamefont
  {Liu}}, \ and\ \bibinfo {author} {\bibfnamefont {James~P.}\ \bibnamefont
  {Sethna}},\ }\bibfield  {title} {\enquote {\bibinfo {title} {Scaling ansatz
  for the jamming transition},}\ }\href {\doibase 10.1073/pnas.1601858113}
  {\bibfield  {journal} {\bibinfo  {journal} {Proceedings of the National
  Academy of Sciences}\ }\textbf {\bibinfo {volume} {113}},\ \bibinfo {pages}
  {9745--9750} (\bibinfo {year} {2016})}\BibitemShut {NoStop}%
\bibitem [{\citenamefont {P. Goodrich}\ \emph {et~al.}(2013)\citenamefont
  {P. Goodrich}, \citenamefont {G. Ellenbroek},\ and\ \citenamefont
  {J. Liu}}]{length_scale_rigidity_2013}%
  \BibitemOpen
  \bibfield  {author} {\bibinfo {author} {\bibfnamefont {Carl}\ \bibnamefont
  {P. Goodrich}}, \bibinfo {author} {\bibfnamefont {Wouter}\ \bibnamefont
  {G. Ellenbroek}}, \ and\ \bibinfo {author} {\bibfnamefont {Andrea}\
  \bibnamefont {J. Liu}},\ }\bibfield  {title} {\enquote {\bibinfo {title}
  {Stability of jammed packings {I}: the rigidity length scale},}\ }\href
  {\doibase 10.1039/C3SM51095F} {\bibfield  {journal} {\bibinfo  {journal}
  {Soft Matter}\ }\textbf {\bibinfo {volume} {9}},\ \bibinfo {pages}
  {10993--10999} (\bibinfo {year} {2013})}\BibitemShut {NoStop}%
\bibitem [{\citenamefont {S. Schoenholz}\ \emph {et~al.}(2013)\citenamefont
  {S. Schoenholz}, \citenamefont {P. Goodrich}, \citenamefont {Kogan},
  \citenamefont {J. Liu},\ and\ \citenamefont
  {R. Nagel}}]{length_scale_transversal_2013}%
  \BibitemOpen
  \bibfield  {author} {\bibinfo {author} {\bibfnamefont {Samuel}\ \bibnamefont
  {S. Schoenholz}}, \bibinfo {author} {\bibfnamefont {Carl}\ \bibnamefont
  {P. Goodrich}}, \bibinfo {author} {\bibfnamefont {Oleg}\ \bibnamefont
  {Kogan}}, \bibinfo {author} {\bibfnamefont {Andrea}\ \bibnamefont {J. Liu}},
  \ and\ \bibinfo {author} {\bibfnamefont {Sidney}\ \bibnamefont {R. Nagel}},\
  }\bibfield  {title} {\enquote {\bibinfo {title} {Stability of jammed packings
  {II}: the transverse length scale},}\ }\href {\doibase 10.1039/C3SM51096D}
  {\bibfield  {journal} {\bibinfo  {journal} {Soft Matter}\ }\textbf {\bibinfo
  {volume} {9}},\ \bibinfo {pages} {11000--11006} (\bibinfo {year}
  {2013})}\BibitemShut {NoStop}%
\bibitem [{\citenamefont {Hexner}\ \emph {et~al.}(2018)\citenamefont {Hexner},
  \citenamefont {Liu},\ and\ \citenamefont
  {Nagel}}]{hexner_two_length_scales_2018}%
  \BibitemOpen
  \bibfield  {author} {\bibinfo {author} {\bibfnamefont {Daniel}\ \bibnamefont
  {Hexner}}, \bibinfo {author} {\bibfnamefont {Andrea~J.}\ \bibnamefont {Liu}},
  \ and\ \bibinfo {author} {\bibfnamefont {Sidney~R.}\ \bibnamefont {Nagel}},\
  }\bibfield  {title} {\enquote {\bibinfo {title} {Two {Diverging} {Length}
  {Scales} in the {Structure} of {Jammed} {Packings}},}\ }\href {\doibase
  10.1103/PhysRevLett.121.115501} {\bibfield  {journal} {\bibinfo  {journal}
  {Physical Review Letters}\ }\textbf {\bibinfo {volume} {121}},\ \bibinfo
  {pages} {115501} (\bibinfo {year} {2018})}\BibitemShut {NoStop}%
\bibitem [{\citenamefont {Hexner}\ \emph {et~al.}(2019)\citenamefont {Hexner},
  \citenamefont {Urbani},\ and\ \citenamefont
  {Zamponi}}]{can_large_packing_2019}%
  \BibitemOpen
  \bibfield  {author} {\bibinfo {author} {\bibfnamefont {Daniel}\ \bibnamefont
  {Hexner}}, \bibinfo {author} {\bibfnamefont {Pierfrancesco}\ \bibnamefont
  {Urbani}}, \ and\ \bibinfo {author} {\bibfnamefont {Francesco}\ \bibnamefont
  {Zamponi}},\ }\bibfield  {title} {\enquote {\bibinfo {title} {Can a {Large}
  {Packing} be {Assembled} from {Smaller} {Ones}?}}\ }\href {\doibase
  10.1103/PhysRevLett.123.068003} {\bibfield  {journal} {\bibinfo  {journal}
  {Physical Review Letters}\ }\textbf {\bibinfo {volume} {123}},\ \bibinfo
  {pages} {068003} (\bibinfo {year} {2019})}\BibitemShut {NoStop}%
\bibitem [{\citenamefont {DeGiuli}\ \emph {et~al.}(2014)\citenamefont
  {DeGiuli}, \citenamefont {Lerner}, \citenamefont {Brito},\ and\ \citenamefont
  {Wyart}}]{degiuli_pnas_2014}%
  \BibitemOpen
  \bibfield  {author} {\bibinfo {author} {\bibfnamefont {Eric}\ \bibnamefont
  {DeGiuli}}, \bibinfo {author} {\bibfnamefont {Edan}\ \bibnamefont {Lerner}},
  \bibinfo {author} {\bibfnamefont {Carolina}\ \bibnamefont {Brito}}, \ and\
  \bibinfo {author} {\bibfnamefont {Matthieu}\ \bibnamefont {Wyart}},\
  }\bibfield  {title} {\enquote {\bibinfo {title} {Force distribution affects
  vibrational properties in hard-sphere glasses},}\ }\href {\doibase
  10.1073/pnas.1415298111} {\bibfield  {journal} {\bibinfo  {journal}
  {Proceedings of the National Academy of Sciences}\ }\textbf {\bibinfo
  {volume} {111}},\ \bibinfo {pages} {17054--17059} (\bibinfo {year}
  {2014})}\BibitemShut {NoStop}%
\bibitem [{\citenamefont {Wyart}(2012)}]{wyart_marginal_stability}%
  \BibitemOpen
  \bibfield  {author} {\bibinfo {author} {\bibfnamefont {Matthieu}\
  \bibnamefont {Wyart}},\ }\bibfield  {title} {\enquote {\bibinfo {title}
  {Marginal {Stability} {Constrains} {Force} and {Pair} {Distributions} at
  {Random} {Close} {Packing}},}\ }\href {\doibase
  10.1103/PhysRevLett.109.125502} {\bibfield  {journal} {\bibinfo  {journal}
  {Physical Review Letters}\ }\textbf {\bibinfo {volume} {109}},\ \bibinfo
  {pages} {125502} (\bibinfo {year} {2012})}\BibitemShut {NoStop}%
\bibitem [{\citenamefont {Müller}\ and\ \citenamefont
  {Wyart}(2015)}]{wyart_muller_review_marginal_2015}%
  \BibitemOpen
  \bibfield  {author} {\bibinfo {author} {\bibfnamefont {Markus}\ \bibnamefont
  {Müller}}\ and\ \bibinfo {author} {\bibfnamefont {Matthieu}\ \bibnamefont
  {Wyart}},\ }\bibfield  {title} {\enquote {\bibinfo {title} {Marginal
  {Stability} in {Structural}, {Spin}, and {Electron} {Glasses}},}\ }\href
  {\doibase 10.1146/annurev-conmatphys-031214-014614} {\bibfield  {journal}
  {\bibinfo  {journal} {Annual Review of Condensed Matter Physics}\ }\textbf
  {\bibinfo {volume} {6}},\ \bibinfo {pages} {177--200} (\bibinfo {year}
  {2015})}\BibitemShut {NoStop}%
\bibitem [{\citenamefont {Berthier}\ \emph {et~al.}(2019)\citenamefont
  {Berthier}, \citenamefont {Biroli}, \citenamefont {Charbonneau},
  \citenamefont {Corwin}, \citenamefont {Franz},\ and\ \citenamefont
  {Zamponi}}]{gardner_perspective}%
  \BibitemOpen
  \bibfield  {author} {\bibinfo {author} {\bibfnamefont {Ludovic}\ \bibnamefont
  {Berthier}}, \bibinfo {author} {\bibfnamefont {Giulio}\ \bibnamefont
  {Biroli}}, \bibinfo {author} {\bibfnamefont {Patrick}\ \bibnamefont
  {Charbonneau}}, \bibinfo {author} {\bibfnamefont {Eric~I.}\ \bibnamefont
  {Corwin}}, \bibinfo {author} {\bibfnamefont {Silvio}\ \bibnamefont {Franz}},
  \ and\ \bibinfo {author} {\bibfnamefont {Francesco}\ \bibnamefont
  {Zamponi}},\ }\bibfield  {title} {\enquote {\bibinfo {title} {Gardner physics
  in amorphous solids and beyond},}\ }\href {\doibase 10.1063/1.5097175}
  {\bibfield  {journal} {\bibinfo  {journal} {The Journal of Chemical Physics}\
  }\textbf {\bibinfo {volume} {151}},\ \bibinfo {pages} {010901} (\bibinfo
  {year} {2019})}\BibitemShut {NoStop}%
\bibitem [{\citenamefont {Franz}\ and\ \citenamefont
  {Parisi}(2016)}]{simplest_jamming}%
  \BibitemOpen
  \bibfield  {author} {\bibinfo {author} {\bibfnamefont {Silvio}\ \bibnamefont
  {Franz}}\ and\ \bibinfo {author} {\bibfnamefont {Giorgio}\ \bibnamefont
  {Parisi}},\ }\bibfield  {title} {\enquote {\bibinfo {title} {The simplest
  model of jamming},}\ }\href {\doibase 10.1088/1751-8113/49/14/145001}
  {\bibfield  {journal} {\bibinfo  {journal} {Journal of Physics A:
  Mathematical and Theoretical}\ }\textbf {\bibinfo {volume} {49}},\ \bibinfo
  {pages} {145001} (\bibinfo {year} {2016})}\BibitemShut {NoStop}%
\bibitem [{\citenamefont {Franz}\ \emph
  {et~al.}(2019{\natexlab{a}})\citenamefont {Franz}, \citenamefont {Sclocchi},\
  and\ \citenamefont {Urbani}}]{franz_critical_perceptron}%
  \BibitemOpen
  \bibfield  {author} {\bibinfo {author} {\bibfnamefont {Silvio}\ \bibnamefont
  {Franz}}, \bibinfo {author} {\bibfnamefont {Antonio}\ \bibnamefont
  {Sclocchi}}, \ and\ \bibinfo {author} {\bibfnamefont {Pierfrancesco}\
  \bibnamefont {Urbani}},\ }\bibfield  {title} {\enquote {\bibinfo {title}
  {Critical jammed phase of the linear perceptron},}\ }\href {\doibase
  10.1103/PhysRevLett.123.115702} {\bibfield  {journal} {\bibinfo  {journal}
  {Physical Review Letters}\ }\textbf {\bibinfo {volume} {123}},\ \bibinfo
  {pages} {115702} (\bibinfo {year} {2019}{\natexlab{a}})}\BibitemShut
  {NoStop}%
\bibitem [{\citenamefont {Spigler}\ \emph {et~al.}(2018)\citenamefont
  {Spigler}, \citenamefont {Geiger}, \citenamefont {d'Ascoli}, \citenamefont
  {Sagun}, \citenamefont {Biroli},\ and\ \citenamefont
  {Wyart}}]{jamming_overparametrization}%
  \BibitemOpen
  \bibfield  {author} {\bibinfo {author} {\bibfnamefont {Stefano}\ \bibnamefont
  {Spigler}}, \bibinfo {author} {\bibfnamefont {Mario}\ \bibnamefont {Geiger}},
  \bibinfo {author} {\bibfnamefont {Stéphane}\ \bibnamefont {d'Ascoli}},
  \bibinfo {author} {\bibfnamefont {Levent}\ \bibnamefont {Sagun}}, \bibinfo
  {author} {\bibfnamefont {Giulio}\ \bibnamefont {Biroli}}, \ and\ \bibinfo
  {author} {\bibfnamefont {Matthieu}\ \bibnamefont {Wyart}},\ }\bibfield
  {title} {\enquote {\bibinfo {title} {A jamming transition from under- to
  over-parametrization affects loss landscape and generalization},}\ }\href
  {http://arxiv.org/abs/1810.09665} {\bibfield  {journal} {\bibinfo  {journal}
  {arXiv:1810.09665 [cond-mat, stat]}\ } (\bibinfo {year} {2018})}\BibitemShut
  {NoStop}%
\bibitem [{\citenamefont {Geiger}\ \emph {et~al.}(2019)\citenamefont {Geiger},
  \citenamefont {Spigler}, \citenamefont {d'Ascoli}, \citenamefont {Sagun},
  \citenamefont {Baity-Jesi}, \citenamefont {Biroli},\ and\ \citenamefont
  {Wyart}}]{jamming_neural_nets}%
  \BibitemOpen
  \bibfield  {author} {\bibinfo {author} {\bibfnamefont {Mario}\ \bibnamefont
  {Geiger}}, \bibinfo {author} {\bibfnamefont {Stefano}\ \bibnamefont
  {Spigler}}, \bibinfo {author} {\bibfnamefont {Stéphane}\ \bibnamefont
  {d'Ascoli}}, \bibinfo {author} {\bibfnamefont {Levent}\ \bibnamefont
  {Sagun}}, \bibinfo {author} {\bibfnamefont {Marco}\ \bibnamefont
  {Baity-Jesi}}, \bibinfo {author} {\bibfnamefont {Giulio}\ \bibnamefont
  {Biroli}}, \ and\ \bibinfo {author} {\bibfnamefont {Matthieu}\ \bibnamefont
  {Wyart}},\ }\bibfield  {title} {\enquote {\bibinfo {title} {Jamming
  transition as a paradigm to understand the loss landscape of deep neural
  networks},}\ }\href {\doibase 10.1103/PhysRevE.100.012115} {\bibfield
  {journal} {\bibinfo  {journal} {Physical Review E}\ }\textbf {\bibinfo
  {volume} {100}},\ \bibinfo {pages} {012115} (\bibinfo {year}
  {2019})}\BibitemShut {NoStop}%
\bibitem [{\citenamefont {Franz}\ \emph
  {et~al.}(2019{\natexlab{b}})\citenamefont {Franz}, \citenamefont {Hwang},\
  and\ \citenamefont {Urbani}}]{jamming_multilayer_supervised}%
  \BibitemOpen
  \bibfield  {author} {\bibinfo {author} {\bibfnamefont {Silvio}\ \bibnamefont
  {Franz}}, \bibinfo {author} {\bibfnamefont {Sungmin}\ \bibnamefont {Hwang}},
  \ and\ \bibinfo {author} {\bibfnamefont {Pierfrancesco}\ \bibnamefont
  {Urbani}},\ }\bibfield  {title} {\enquote {\bibinfo {title} {Jamming in
  {{Multilayer Supervised Learning Models}}},}\ }\href {\doibase
  10.1103/PhysRevLett.123.160602} {\bibfield  {journal} {\bibinfo  {journal}
  {Physical Review Letters}\ }\textbf {\bibinfo {volume} {123}},\ \bibinfo
  {pages} {160602} (\bibinfo {year} {2019}{\natexlab{b}})}\BibitemShut
  {NoStop}%
\bibitem [{\citenamefont {Antenucci}\ \emph {et~al.}(2019)\citenamefont
  {Antenucci}, \citenamefont {Franz}, \citenamefont {Urbani},\ and\
  \citenamefont {Zdeborov\'{a}}}]{inference_glassy_nature}%
  \BibitemOpen
  \bibfield  {author} {\bibinfo {author} {\bibfnamefont {Fabrizio}\
  \bibnamefont {Antenucci}}, \bibinfo {author} {\bibfnamefont {Silvio}\
  \bibnamefont {Franz}}, \bibinfo {author} {\bibfnamefont {Pierfrancesco}\
  \bibnamefont {Urbani}}, \ and\ \bibinfo {author} {\bibfnamefont {Lenka}\
  \bibnamefont {Zdeborov\'{a}}},\ }\bibfield  {title} {\enquote {\bibinfo
  {title} {Glassy {Nature} of the {Hard} {Phase} in {Inference} {Problems}},}\
  }\href {\doibase 10.1103/PhysRevX.9.011020} {\bibfield  {journal} {\bibinfo
  {journal} {Physical Review X}\ }\textbf {\bibinfo {volume} {9}},\ \bibinfo
  {pages} {011020} (\bibinfo {year} {2019})}\BibitemShut {NoStop}%
\bibitem [{\citenamefont {Franz}\ \emph {et~al.}(2017)\citenamefont {Franz},
  \citenamefont {Parisi}, \citenamefont {Sevelev}, \citenamefont {Urbani},\
  and\ \citenamefont {Zamponi}}]{jamming_sat_unsat}%
  \BibitemOpen
  \bibfield  {author} {\bibinfo {author} {\bibfnamefont {Silvio}\ \bibnamefont
  {Franz}}, \bibinfo {author} {\bibfnamefont {Giorgio}\ \bibnamefont {Parisi}},
  \bibinfo {author} {\bibfnamefont {Maxim}\ \bibnamefont {Sevelev}}, \bibinfo
  {author} {\bibfnamefont {Pierfrancesco}\ \bibnamefont {Urbani}}, \ and\
  \bibinfo {author} {\bibfnamefont {Francesco}\ \bibnamefont {Zamponi}},\
  }\bibfield  {title} {\enquote {\bibinfo {title} {Universality of the
  {SAT}-{UNSAT} (jamming) threshold in non-convex continuous constraint
  satisfaction problems},}\ }\href {\doibase 10.21468/SciPostPhys.2.3.019}
  {\bibfield  {journal} {\bibinfo  {journal} {SciPost Physics}\ }\textbf
  {\bibinfo {volume} {2}},\ \bibinfo {pages} {019} (\bibinfo {year}
  {2017})}\BibitemShut {NoStop}%
\bibitem [{\citenamefont {Krzakala}\ and\ \citenamefont
  {Kurchan}(2007)}]{krzakala_kurchan_landscape_2007}%
  \BibitemOpen
  \bibfield  {author} {\bibinfo {author} {\bibfnamefont {Florent}\ \bibnamefont
  {Krzakala}}\ and\ \bibinfo {author} {\bibfnamefont {Jorge}\ \bibnamefont
  {Kurchan}},\ }\bibfield  {title} {\enquote {\bibinfo {title} {Landscape
  analysis of constraint satisfaction problems},}\ }\href {\doibase
  10.1103/PhysRevE.76.021122} {\bibfield  {journal} {\bibinfo  {journal}
  {Physical Review E}\ }\textbf {\bibinfo {volume} {76}},\ \bibinfo {pages}
  {021122} (\bibinfo {year} {2007})}\BibitemShut {NoStop}%
\bibitem [{\citenamefont {Franz}\ \emph {et~al.}(2020)\citenamefont {Franz},
  \citenamefont {Sclocchi},\ and\ \citenamefont
  {Urbani}}]{franz_linear_spheres}%
  \BibitemOpen
  \bibfield  {author} {\bibinfo {author} {\bibfnamefont {Silvio}\ \bibnamefont
  {Franz}}, \bibinfo {author} {\bibfnamefont {Antonio}\ \bibnamefont
  {Sclocchi}}, \ and\ \bibinfo {author} {\bibfnamefont {Pierfrancesco}\
  \bibnamefont {Urbani}},\ }\bibfield  {title} {\enquote {\bibinfo {title}
  {Critical energy landscape of linear soft spheres},}\ }\href {\doibase
  10.21468/SciPostPhys.9.1.012} {\bibfield  {journal} {\bibinfo  {journal}
  {SciPost Physics}\ }\textbf {\bibinfo {volume} {9}},\ \bibinfo {pages} {012}
  (\bibinfo {year} {2020})}\BibitemShut {NoStop}%
\bibitem [{\citenamefont {Charbonneau}\ \emph {et~al.}(2019)\citenamefont
  {Charbonneau}, \citenamefont {Corwin}, \citenamefont {Fu}, \citenamefont
  {Tsekenis},\ and\ \citenamefont {van~der Naald}}]{gardner-crystals}%
  \BibitemOpen
  \bibfield  {author} {\bibinfo {author} {\bibfnamefont {Patrick}\ \bibnamefont
  {Charbonneau}}, \bibinfo {author} {\bibfnamefont {Eric~I.}\ \bibnamefont
  {Corwin}}, \bibinfo {author} {\bibfnamefont {Lin}\ \bibnamefont {Fu}},
  \bibinfo {author} {\bibfnamefont {Georgios}\ \bibnamefont {Tsekenis}}, \ and\
  \bibinfo {author} {\bibfnamefont {Michael}\ \bibnamefont {van~der Naald}},\
  }\bibfield  {title} {\enquote {\bibinfo {title} {Glassy, {Gardner}-like
  phenomenology in minimally polydisperse crystalline systems},}\ }\href
  {\doibase 10.1103/PhysRevE.99.020901} {\bibfield  {journal} {\bibinfo
  {journal} {Physical Review E}\ }\textbf {\bibinfo {volume} {99}},\ \bibinfo
  {pages} {020901(R)} (\bibinfo {year} {2019})}\BibitemShut {NoStop}%
\bibitem [{\citenamefont {Tsekenis}(2020)}]{tsekenis_jamming_2020}%
  \BibitemOpen
  \bibfield  {author} {\bibinfo {author} {\bibfnamefont {Georgios}\
  \bibnamefont {Tsekenis}},\ }\bibfield  {title} {\enquote {\bibinfo {title}
  {Jamming {Criticality} of {Near}-{Crystals}},}\ }\href
  {http://arxiv.org/abs/2006.07373} {\bibfield  {journal} {\bibinfo  {journal}
  {arXiv:2006.07373 [cond-mat]}\ } (\bibinfo {year} {2020})}\BibitemShut
  {NoStop}%
\bibitem [{\citenamefont {Ikeda}\ \emph {et~al.}(2020)\citenamefont {Ikeda},
  \citenamefont {Brito},\ and\ \citenamefont
  {Wyart}}]{ikeda_infinitesimal_2020}%
  \BibitemOpen
  \bibfield  {author} {\bibinfo {author} {\bibfnamefont {Harukuni}\
  \bibnamefont {Ikeda}}, \bibinfo {author} {\bibfnamefont {Carolina}\
  \bibnamefont {Brito}}, \ and\ \bibinfo {author} {\bibfnamefont {Matthieu}\
  \bibnamefont {Wyart}},\ }\bibfield  {title} {\enquote {\bibinfo {title}
  {Infinitesimal asphericity changes the universality of the jamming
  transition},}\ }\href {\doibase 10.1088/1742-5468/ab74cb} {\bibfield
  {journal} {\bibinfo  {journal} {Journal of Statistical Mechanics: Theory and
  Experiment}\ }\textbf {\bibinfo {volume} {2020}},\ \bibinfo {pages} {033302}
  (\bibinfo {year} {2020})}\BibitemShut {NoStop}%
\bibitem [{\citenamefont {Brito}\ \emph {et~al.}(2018)\citenamefont {Brito},
  \citenamefont {Ikeda}, \citenamefont {Urbani}, \citenamefont {Wyart},\ and\
  \citenamefont {Zamponi}}]{brito_universality_2018}%
  \BibitemOpen
  \bibfield  {author} {\bibinfo {author} {\bibfnamefont {Carolina}\
  \bibnamefont {Brito}}, \bibinfo {author} {\bibfnamefont {Harukuni}\
  \bibnamefont {Ikeda}}, \bibinfo {author} {\bibfnamefont {Pierfrancesco}\
  \bibnamefont {Urbani}}, \bibinfo {author} {\bibfnamefont {Matthieu}\
  \bibnamefont {Wyart}}, \ and\ \bibinfo {author} {\bibfnamefont {Francesco}\
  \bibnamefont {Zamponi}},\ }\bibfield  {title} {\enquote {\bibinfo {title}
  {Universality of jamming of nonspherical particles},}\ }\href {\doibase
  10.1073/pnas.1812457115} {\bibfield  {journal} {\bibinfo  {journal}
  {Proceedings of the National Academy of Sciences}\ }\textbf {\bibinfo
  {volume} {115}},\ \bibinfo {pages} {11736--11741} (\bibinfo {year}
  {2018})}\BibitemShut {NoStop}%
\bibitem [{\citenamefont {Kallus}(2016)}]{perceptron_size_scaling}%
  \BibitemOpen
  \bibfield  {author} {\bibinfo {author} {\bibfnamefont {Yoav}\ \bibnamefont
  {Kallus}},\ }\bibfield  {title} {\enquote {\bibinfo {title} {Scaling collapse
  at the jamming transition},}\ }\href {\doibase 10.1103/PhysRevE.93.012902}
  {\bibfield  {journal} {\bibinfo  {journal} {Physical Review E}\ }\textbf
  {\bibinfo {volume} {93}},\ \bibinfo {pages} {012902} (\bibinfo {year}
  {2016})}\BibitemShut {NoStop}%
\bibitem [{\citenamefont {Mari}\ and\ \citenamefont
  {Kurchan}(2011)}]{mk_definition_2011}%
  \BibitemOpen
  \bibfield  {author} {\bibinfo {author} {\bibfnamefont {Romain}\ \bibnamefont
  {Mari}}\ and\ \bibinfo {author} {\bibfnamefont {Jorge}\ \bibnamefont
  {Kurchan}},\ }\bibfield  {title} {\enquote {\bibinfo {title} {Dynamical
  transition of glasses: {From} exact to approximate},}\ }\href {\doibase
  10.1063/1.3626802} {\bibfield  {journal} {\bibinfo  {journal} {The Journal of
  Chemical Physics}\ }\textbf {\bibinfo {volume} {135}},\ \bibinfo {pages}
  {124504} (\bibinfo {year} {2011})}\BibitemShut {NoStop}%
\bibitem [{\citenamefont {O'Hern}\ \emph {et~al.}(2002)\citenamefont {O'Hern},
  \citenamefont {Langer}, \citenamefont {Liu},\ and\ \citenamefont
  {Nagel}}]{ohern_random_2002}%
  \BibitemOpen
  \bibfield  {author} {\bibinfo {author} {\bibfnamefont {Corey~S.}\
  \bibnamefont {O'Hern}}, \bibinfo {author} {\bibfnamefont {Stephen~A.}\
  \bibnamefont {Langer}}, \bibinfo {author} {\bibfnamefont {Andrea~J.}\
  \bibnamefont {Liu}}, \ and\ \bibinfo {author} {\bibfnamefont {Sidney~R.}\
  \bibnamefont {Nagel}},\ }\bibfield  {title} {\enquote {\bibinfo {title}
  {Random {{Packings}} of {{Frictionless Particles}}},}\ }\href {\doibase
  10.1103/PhysRevLett.88.075507} {\bibfield  {journal} {\bibinfo  {journal}
  {Physical Review Letters}\ }\textbf {\bibinfo {volume} {88}},\ \bibinfo
  {pages} {075507} (\bibinfo {year} {2002})}\BibitemShut {NoStop}%
\bibitem [{\citenamefont {Hagh}\ \emph {et~al.}(2019)\citenamefont {Hagh},
  \citenamefont {Corwin}, \citenamefont {Stephenson},\ and\ \citenamefont
  {Thorpe}}]{broader_view_jamming_2019}%
  \BibitemOpen
  \bibfield  {author} {\bibinfo {author} {\bibfnamefont {Varda~F.}\
  \bibnamefont {Hagh}}, \bibinfo {author} {\bibfnamefont {Eric~I.}\
  \bibnamefont {Corwin}}, \bibinfo {author} {\bibfnamefont {Kenneth}\
  \bibnamefont {Stephenson}}, \ and\ \bibinfo {author} {\bibfnamefont {M.~F.}\
  \bibnamefont {Thorpe}},\ }\bibfield  {title} {\enquote {\bibinfo {title} {A
  broader view on jamming: From spring networks to circle packings},}\ }\href
  {\doibase 10.1039/C8SM01768A} {\bibfield  {journal} {\bibinfo  {journal}
  {Soft Matter}\ }\textbf {\bibinfo {volume} {15}},\ \bibinfo {pages}
  {3076--3084} (\bibinfo {year} {2019})}\BibitemShut {NoStop}%
\bibitem [{\citenamefont {Donev}\ \emph {et~al.}(2005)\citenamefont {Donev},
  \citenamefont {Torquato},\ and\ \citenamefont
  {Stillinger}}]{donev_pair_2005}%
  \BibitemOpen
  \bibfield  {author} {\bibinfo {author} {\bibfnamefont {Aleksandar}\
  \bibnamefont {Donev}}, \bibinfo {author} {\bibfnamefont {Salvatore}\
  \bibnamefont {Torquato}}, \ and\ \bibinfo {author} {\bibfnamefont {Frank~H.}\
  \bibnamefont {Stillinger}},\ }\bibfield  {title} {\enquote {\bibinfo {title}
  {Pair correlation function characteristics of nearly jammed disordered and
  ordered hard-sphere packings},}\ }\href {\doibase 10.1103/PhysRevE.71.011105}
  {\bibfield  {journal} {\bibinfo  {journal} {Physical Review E}\ }\textbf
  {\bibinfo {volume} {71}},\ \bibinfo {pages} {011105} (\bibinfo {year}
  {2005})}\BibitemShut {NoStop}%
\bibitem [{\citenamefont {Bitzek}\ \emph {et~al.}(2006)\citenamefont {Bitzek},
  \citenamefont {Koskinen}, \citenamefont {G\"{a}hler}, \citenamefont
  {Moseler},\ and\ \citenamefont {Gumbsch}}]{FIRE}%
  \BibitemOpen
  \bibfield  {author} {\bibinfo {author} {\bibfnamefont {Erik}\ \bibnamefont
  {Bitzek}}, \bibinfo {author} {\bibfnamefont {Pekka}\ \bibnamefont
  {Koskinen}}, \bibinfo {author} {\bibfnamefont {Franz}\ \bibnamefont
  {G\"{a}hler}}, \bibinfo {author} {\bibfnamefont {Michael}\ \bibnamefont
  {Moseler}}, \ and\ \bibinfo {author} {\bibfnamefont {Peter}\ \bibnamefont
  {Gumbsch}},\ }\bibfield  {title} {\enquote {\bibinfo {title} {Structural
  {Relaxation} {Made} {Simple}},}\ }\href {\doibase
  10.1103/PhysRevLett.97.170201} {\bibfield  {journal} {\bibinfo  {journal}
  {Physical Review Letters}\ }\textbf {\bibinfo {volume} {97}},\ \bibinfo
  {pages} {170201} (\bibinfo {year} {2006})}\BibitemShut {NoStop}%
\bibitem [{\citenamefont {Morse}\ and\ \citenamefont
  {Corwin}(2014)}]{morse2014geometric}%
  \BibitemOpen
  \bibfield  {author} {\bibinfo {author} {\bibfnamefont {Peter~K.}\
  \bibnamefont {Morse}}\ and\ \bibinfo {author} {\bibfnamefont {Eric~I.}\
  \bibnamefont {Corwin}},\ }\bibfield  {title} {\enquote {\bibinfo {title}
  {Geometric {{Signatures}} of {{Jamming}} in the {{Mechanical Vacuum}}},}\
  }\href {\doibase 10.1103/PhysRevLett.112.115701} {\bibfield  {journal}
  {\bibinfo  {journal} {Physical Review Letters}\ }\textbf {\bibinfo {volume}
  {112}},\ \bibinfo {pages} {115701} (\bibinfo {year} {2014})}\BibitemShut
  {NoStop}%
\bibitem [{\citenamefont {Charbonneau}\ \emph {et~al.}(2016)\citenamefont
  {Charbonneau}, \citenamefont {Corwin}, \citenamefont {Parisi}, \citenamefont
  {Poncet},\ and\ \citenamefont {Zamponi}}]{charbonneau2016universal}%
  \BibitemOpen
  \bibfield  {author} {\bibinfo {author} {\bibfnamefont {Patrick}\ \bibnamefont
  {Charbonneau}}, \bibinfo {author} {\bibfnamefont {Eric~I.}\ \bibnamefont
  {Corwin}}, \bibinfo {author} {\bibfnamefont {Giorgio}\ \bibnamefont
  {Parisi}}, \bibinfo {author} {\bibfnamefont {Alexis}\ \bibnamefont {Poncet}},
  \ and\ \bibinfo {author} {\bibfnamefont {Francesco}\ \bibnamefont
  {Zamponi}},\ }\bibfield  {title} {\enquote {\bibinfo {title} {Universal
  {{Non}}-{{Debye Scaling}} in the {{Density}} of {{States}} of {{Amorphous
  Solids}}},}\ }\href {\doibase 10.1103/PhysRevLett.117.045503} {\bibfield
  {journal} {\bibinfo  {journal} {Physical Review Letters}\ }\textbf {\bibinfo
  {volume} {117}},\ \bibinfo {pages} {045503} (\bibinfo {year}
  {2016})}\BibitemShut {NoStop}%
\bibitem [{\citenamefont {Morse}\ and\ \citenamefont
  {Corwin}(2017)}]{morse2017echoes}%
  \BibitemOpen
  \bibfield  {author} {\bibinfo {author} {\bibfnamefont {Peter~K.}\
  \bibnamefont {Morse}}\ and\ \bibinfo {author} {\bibfnamefont {Eric~I.}\
  \bibnamefont {Corwin}},\ }\bibfield  {title} {\enquote {\bibinfo {title}
  {Echoes of the {{Glass Transition}} in {{Athermal Soft Spheres}}},}\ }\href
  {\doibase 10.1103/PhysRevLett.119.118003} {\bibfield  {journal} {\bibinfo
  {journal} {Physical Review Letters}\ }\textbf {\bibinfo {volume} {119}},\
  \bibinfo {pages} {118003} (\bibinfo {year} {2017})}\BibitemShut {NoStop}%
\bibitem [{\citenamefont {Artiaco}\ \emph {et~al.}(2020)\citenamefont
  {Artiaco}, \citenamefont {Baldan},\ and\ \citenamefont
  {Parisi}}]{artiaco_baldan_2020}%
  \BibitemOpen
  \bibfield  {author} {\bibinfo {author} {\bibfnamefont {Claudia}\ \bibnamefont
  {Artiaco}}, \bibinfo {author} {\bibfnamefont {Paolo}\ \bibnamefont {Baldan}},
  \ and\ \bibinfo {author} {\bibfnamefont {Giorgio}\ \bibnamefont {Parisi}},\
  }\bibfield  {title} {\enquote {\bibinfo {title} {Exploratory study of the
  glassy landscape near jamming},}\ }\href {\doibase
  10.1103/PhysRevE.101.052605} {\bibfield  {journal} {\bibinfo  {journal}
  {Physical Review E}\ }\textbf {\bibinfo {volume} {101}},\ \bibinfo {pages}
  {052605} (\bibinfo {year} {2020})}\BibitemShut {NoStop}%
\bibitem [{\citenamefont {D{\'i}az Hern{\'a}ndez~Rojas}\ \emph
  {et~al.}(2021)\citenamefont {D{\'i}az Hern{\'a}ndez~Rojas}, \citenamefont
  {Parisi},\ and\ \citenamefont
  {{Ricci-Tersenghi}}}]{diaz_hernandez_rojas_inferring_2020}%
  \BibitemOpen
  \bibfield  {author} {\bibinfo {author} {\bibfnamefont {Rafael}\ \bibnamefont
  {D{\'i}az Hern{\'a}ndez~Rojas}}, \bibinfo {author} {\bibfnamefont {Giorgio}\
  \bibnamefont {Parisi}}, \ and\ \bibinfo {author} {\bibfnamefont {Federico}\
  \bibnamefont {{Ricci-Tersenghi}}},\ }\bibfield  {title} {\enquote {\bibinfo
  {title} {Inferring the particle-wise dynamics of amorphous solids from the
  local structure at the jamming point},}\ }\href {\doibase 10.1039/C9SM02283J}
  {\bibfield  {journal} {\bibinfo  {journal} {Soft Matter}\ }\textbf {\bibinfo
  {volume} {17}},\ \bibinfo {pages} {1056--1083} (\bibinfo {year}
  {2021})}\BibitemShut {NoStop}%
\bibitem [{\citenamefont {Charbonneau}\ and\ \citenamefont
  {Morse}(2021)}]{charbonneauMemoryFormationJammed2021}%
  \BibitemOpen
  \bibfield  {author} {\bibinfo {author} {\bibfnamefont {Patrick}\ \bibnamefont
  {Charbonneau}}\ and\ \bibinfo {author} {\bibfnamefont {Peter~K.}\
  \bibnamefont {Morse}},\ }\bibfield  {title} {\enquote {\bibinfo {title}
  {Memory {{Formation}} in {{Jammed Hard Spheres}}},}\ }\href {\doibase
  10.1103/PhysRevLett.126.088001} {\bibfield  {journal} {\bibinfo  {journal}
  {Physical Review Letters}\ }\textbf {\bibinfo {volume} {126}},\ \bibinfo
  {pages} {088001} (\bibinfo {year} {2021})}\BibitemShut {NoStop}%
\bibitem [{\citenamefont {Frenkel}(2015)}]{frenkelOrderEntropy2015}%
  \BibitemOpen
  \bibfield  {author} {\bibinfo {author} {\bibfnamefont {Daan}\ \bibnamefont
  {Frenkel}},\ }\bibfield  {title} {\enquote {\bibinfo {title} {Order through
  entropy},}\ }\href {\doibase 10.1038/nmat4178} {\bibfield  {journal}
  {\bibinfo  {journal} {Nature Materials}\ }\textbf {\bibinfo {volume} {14}},\
  \bibinfo {pages} {9--12} (\bibinfo {year} {2015})}\BibitemShut {NoStop}%
\bibitem [{\citenamefont {Xu}\ \emph {et~al.}(2005)\citenamefont {Xu},
  \citenamefont {Blawzdziewicz},\ and\ \citenamefont
  {O'Hern}}]{xuRandomClosePacking2005}%
  \BibitemOpen
  \bibfield  {author} {\bibinfo {author} {\bibfnamefont {Ning}\ \bibnamefont
  {Xu}}, \bibinfo {author} {\bibfnamefont {Jerzy}\ \bibnamefont
  {Blawzdziewicz}}, \ and\ \bibinfo {author} {\bibfnamefont {Corey~S.}\
  \bibnamefont {O'Hern}},\ }\bibfield  {title} {\enquote {\bibinfo {title}
  {Random close packing revisited: {{Ways}} to pack frictionless disks},}\
  }\href {\doibase 10.1103/PhysRevE.71.061306} {\bibfield  {journal} {\bibinfo
  {journal} {Physical Review E}\ }\textbf {\bibinfo {volume} {71}},\ \bibinfo
  {pages} {061306} (\bibinfo {year} {2005})}\BibitemShut {NoStop}%
\bibitem [{\citenamefont {Arceri}\ and\ \citenamefont
  {Corwin}(2020)}]{arceriVibrationalPropertiesHard2020}%
  \BibitemOpen
  \bibfield  {author} {\bibinfo {author} {\bibfnamefont {Francesco}\
  \bibnamefont {Arceri}}\ and\ \bibinfo {author} {\bibfnamefont {Eric~I.}\
  \bibnamefont {Corwin}},\ }\bibfield  {title} {\enquote {\bibinfo {title}
  {Vibrational {{Properties}} of {{Hard}} and {{Soft Spheres Are Unified}} at
  {{Jamming}}},}\ }\href {\doibase 10.1103/PhysRevLett.124.238002} {\bibfield
  {journal} {\bibinfo  {journal} {Physical Review Letters}\ }\textbf {\bibinfo
  {volume} {124}},\ \bibinfo {pages} {238002} (\bibinfo {year}
  {2020})}\BibitemShut {NoStop}%
\bibitem [{\citenamefont {Charbonneau}\ \emph
  {et~al.}(2014{\natexlab{c}})\citenamefont {Charbonneau}, \citenamefont {Jin},
  \citenamefont {Parisi},\ and\ \citenamefont {Zamponi}}]{mk_hopping_2014}%
  \BibitemOpen
  \bibfield  {author} {\bibinfo {author} {\bibfnamefont {Patrick}\ \bibnamefont
  {Charbonneau}}, \bibinfo {author} {\bibfnamefont {Yuliang}\ \bibnamefont
  {Jin}}, \bibinfo {author} {\bibfnamefont {Giorgio}\ \bibnamefont {Parisi}}, \
  and\ \bibinfo {author} {\bibfnamefont {Francesco}\ \bibnamefont {Zamponi}},\
  }\bibfield  {title} {\enquote {\bibinfo {title} {Hopping and the
  {Stokes}–{Einstein} relation breakdown in simple glass formers},}\ }\href
  {\doibase 10.1073/pnas.1417182111} {\bibfield  {journal} {\bibinfo  {journal}
  {Proceedings of the National Academy of Sciences}\ }\textbf {\bibinfo
  {volume} {111}},\ \bibinfo {pages} {15025--15030} (\bibinfo {year}
  {2014}{\natexlab{c}})}\BibitemShut {NoStop}%
\bibitem [{\citenamefont {Charbonneau}\ \emph
  {et~al.}(2015{\natexlab{b}})\citenamefont {Charbonneau}, \citenamefont {Jin},
  \citenamefont {Parisi}, \citenamefont {Rainone}, \citenamefont {Seoane},\
  and\ \citenamefont {Zamponi}}]{mk_gardner_2015}%
  \BibitemOpen
  \bibfield  {author} {\bibinfo {author} {\bibfnamefont {Patrick}\ \bibnamefont
  {Charbonneau}}, \bibinfo {author} {\bibfnamefont {Yuliang}\ \bibnamefont
  {Jin}}, \bibinfo {author} {\bibfnamefont {Giorgio}\ \bibnamefont {Parisi}},
  \bibinfo {author} {\bibfnamefont {Corrado}\ \bibnamefont {Rainone}}, \bibinfo
  {author} {\bibfnamefont {Beatriz}\ \bibnamefont {Seoane}}, \ and\ \bibinfo
  {author} {\bibfnamefont {Francesco}\ \bibnamefont {Zamponi}},\ }\bibfield
  {title} {\enquote {\bibinfo {title} {Numerical detection of the {Gardner}
  transition in a mean-field glass former},}\ }\href {\doibase
  10.1103/PhysRevE.92.012316} {\bibfield  {journal} {\bibinfo  {journal}
  {Physical Review E}\ }\textbf {\bibinfo {volume} {92}},\ \bibinfo {pages}
  {012316} (\bibinfo {year} {2015}{\natexlab{b}})}\BibitemShut {NoStop}%
\bibitem [{\citenamefont {Newman}(2005)}]{newman_power_laws}%
  \BibitemOpen
  \bibfield  {author} {\bibinfo {author} {\bibfnamefont {M.~E.~J.}\
  \bibnamefont {Newman}},\ }\bibfield  {title} {\enquote {\bibinfo {title}
  {Power laws, {Pareto} distributions and {Zipf}'s law},}\ }\href {\doibase
  10.1080/00107510500052444} {\bibfield  {journal} {\bibinfo  {journal}
  {Contemporary Physics}\ }\textbf {\bibinfo {volume} {46}},\ \bibinfo {pages}
  {323--351} (\bibinfo {year} {2005})}\BibitemShut {NoStop}%
\bibitem [{\citenamefont {Amit}\ and\ \citenamefont
  {Martin-Mayor}(2005)}]{amit2005field}%
  \BibitemOpen
  \bibfield  {author} {\bibinfo {author} {\bibfnamefont {Daniel~J}\
  \bibnamefont {Amit}}\ and\ \bibinfo {author} {\bibfnamefont {Victor}\
  \bibnamefont {Martin-Mayor}},\ }\href@noop {} {\emph {\bibinfo {title} {Field
  theory, the renormalization group, and critical phenomena: graphs to
  computers}}}\ (\bibinfo  {publisher} {World Scientific Publishing Company},\
  \bibinfo {year} {2005})\BibitemShut {NoStop}%
\bibitem [{\citenamefont {Newman}\ and\ \citenamefont
  {Barkema}(1999)}]{MC-book}%
  \BibitemOpen
  \bibfield  {author} {\bibinfo {author} {\bibfnamefont {M}~\bibnamefont
  {Newman}}\ and\ \bibinfo {author} {\bibfnamefont {G}~\bibnamefont
  {Barkema}},\ }\href@noop {} {\emph {\bibinfo {title} {{Monte Carlo} methods
  in statistical physics}}}\ (\bibinfo  {publisher} {Oxford University Press:
  New York, USA},\ \bibinfo {year} {1999})\BibitemShut {NoStop}%
\bibitem [{\citenamefont {Wang}\ and\ \citenamefont
  {Young}(1993)}]{wang1993monte}%
  \BibitemOpen
  \bibfield  {author} {\bibinfo {author} {\bibfnamefont {J.}~\bibnamefont
  {Wang}}\ and\ \bibinfo {author} {\bibfnamefont {A.~P.}\ \bibnamefont
  {Young}},\ }\bibfield  {title} {\enquote {\bibinfo {title} {Monte {{Carlo}}
  study of the six-dimensional {{Ising}} spin glass},}\ }\href {\doibase
  10.1088/0305-4470/26/5/025} {\bibfield  {journal} {\bibinfo  {journal}
  {Journal of Physics A: Mathematical and General}\ }\textbf {\bibinfo {volume}
  {26}},\ \bibinfo {pages} {1063} (\bibinfo {year} {1993})}\BibitemShut
  {NoStop}%
\bibitem [{\citenamefont {{Ruiz-Lorenzo}}(1998)}]{ruiz1998logarithmic}%
  \BibitemOpen
  \bibfield  {author} {\bibinfo {author} {\bibfnamefont {Juan~J.}\ \bibnamefont
  {{Ruiz-Lorenzo}}},\ }\bibfield  {title} {\enquote {\bibinfo {title}
  {Logarithmic corrections for spin glasses, percolation and {{Lee}}-{{Yang}}
  singularities in six dimensions},}\ }\href {\doibase
  10.1088/0305-4470/31/44/006} {\bibfield  {journal} {\bibinfo  {journal}
  {Journal of Physics A: Mathematical and General}\ }\textbf {\bibinfo {volume}
  {31}},\ \bibinfo {pages} {8773--8787} (\bibinfo {year} {1998})}\BibitemShut
  {NoStop}%
\bibitem [{\citenamefont {Kenna}(2004)}]{Kenna2004}%
  \BibitemOpen
  \bibfield  {author} {\bibinfo {author} {\bibfnamefont {R}~\bibnamefont
  {Kenna}},\ }\bibfield  {title} {\enquote {\bibinfo {title} {Finite size
  scaling for {{O}}({{N}}) {$\Phi$}4-theory at the upper critical dimension},}\
  }\href {\doibase 10.1016/j.nuclphysb.2004.05.012} {\bibfield  {journal}
  {\bibinfo  {journal} {Nuclear Physics B}\ }\textbf {\bibinfo {volume}
  {691}},\ \bibinfo {pages} {292--304} (\bibinfo {year} {2004})}\BibitemShut
  {NoStop}%
\bibitem [{\citenamefont {Ikeda}\ \emph {et~al.}(2019)\citenamefont {Ikeda},
  \citenamefont {Urbani},\ and\ \citenamefont {Zamponi}}]{MF_non_spheres}%
  \BibitemOpen
  \bibfield  {author} {\bibinfo {author} {\bibfnamefont {Harukuni}\
  \bibnamefont {Ikeda}}, \bibinfo {author} {\bibfnamefont {Pierfrancesco}\
  \bibnamefont {Urbani}}, \ and\ \bibinfo {author} {\bibfnamefont {Francesco}\
  \bibnamefont {Zamponi}},\ }\bibfield  {title} {\enquote {\bibinfo {title}
  {Mean field theory of jamming of nonspherical particles},}\ }\href {\doibase
  10.1088/1751-8121/ab3079} {\bibfield  {journal} {\bibinfo  {journal} {Journal
  of Physics A: Mathematical and Theoretical}\ }\textbf {\bibinfo {volume}
  {52}},\ \bibinfo {pages} {344001} (\bibinfo {year} {2019})}\BibitemShut
  {NoStop}%
\bibitem [{\citenamefont {Lucibello}\ \emph
  {et~al.}(2014{\natexlab{a}})\citenamefont {Lucibello}, \citenamefont
  {Morone}, \citenamefont {Parisi}, \citenamefont {Ricci-Tersenghi},\ and\
  \citenamefont {Rizzo}}]{lucibello_anomalous_2014}%
  \BibitemOpen
  \bibfield  {author} {\bibinfo {author} {\bibfnamefont {C}~\bibnamefont
  {Lucibello}}, \bibinfo {author} {\bibfnamefont {F}~\bibnamefont {Morone}},
  \bibinfo {author} {\bibfnamefont {G}~\bibnamefont {Parisi}}, \bibinfo
  {author} {\bibfnamefont {F}~\bibnamefont {Ricci-Tersenghi}}, \ and\ \bibinfo
  {author} {\bibfnamefont {Tommaso}\ \bibnamefont {Rizzo}},\ }\bibfield
  {title} {\enquote {\bibinfo {title} {Anomalous finite size corrections in
  random field models},}\ }\href {\doibase 10.1088/1742-5468/2014/10/P10025}
  {\bibfield  {journal} {\bibinfo  {journal} {Journal of Statistical Mechanics:
  Theory and Experiment}\ }\textbf {\bibinfo {volume} {2014}},\ \bibinfo
  {pages} {P10025} (\bibinfo {year} {2014}{\natexlab{a}})}\BibitemShut
  {NoStop}%
\bibitem [{\citenamefont {Lucibello}\ \emph
  {et~al.}(2014{\natexlab{b}})\citenamefont {Lucibello}, \citenamefont
  {Morone}, \citenamefont {Parisi}, \citenamefont {Ricci-Tersenghi},\ and\
  \citenamefont {Rizzo}}]{lucibello_finite-size_2014}%
  \BibitemOpen
  \bibfield  {author} {\bibinfo {author} {\bibfnamefont {C.}~\bibnamefont
  {Lucibello}}, \bibinfo {author} {\bibfnamefont {F.}~\bibnamefont {Morone}},
  \bibinfo {author} {\bibfnamefont {G.}~\bibnamefont {Parisi}}, \bibinfo
  {author} {\bibfnamefont {F.}~\bibnamefont {Ricci-Tersenghi}}, \ and\ \bibinfo
  {author} {\bibfnamefont {Tommaso}\ \bibnamefont {Rizzo}},\ }\bibfield
  {title} {\enquote {\bibinfo {title} {Finite-size corrections to disordered
  {Ising} models on random regular graphs},}\ }\href {\doibase
  10.1103/PhysRevE.90.012146} {\bibfield  {journal} {\bibinfo  {journal}
  {Physical Review E}\ }\textbf {\bibinfo {volume} {90}},\ \bibinfo {pages}
  {012146} (\bibinfo {year} {2014}{\natexlab{b}})}\BibitemShut {NoStop}%
\bibitem [{\citenamefont {Ferrari}\ \emph {et~al.}(2013)\citenamefont
  {Ferrari}, \citenamefont {Lucibello}, \citenamefont {Morone}, \citenamefont
  {Parisi}, \citenamefont {Ricci-Tersenghi},\ and\ \citenamefont
  {Rizzo}}]{ferrari_finite-size_2013}%
  \BibitemOpen
  \bibfield  {author} {\bibinfo {author} {\bibfnamefont {U.}~\bibnamefont
  {Ferrari}}, \bibinfo {author} {\bibfnamefont {C.}~\bibnamefont {Lucibello}},
  \bibinfo {author} {\bibfnamefont {F.}~\bibnamefont {Morone}}, \bibinfo
  {author} {\bibfnamefont {G.}~\bibnamefont {Parisi}}, \bibinfo {author}
  {\bibfnamefont {F.}~\bibnamefont {Ricci-Tersenghi}}, \ and\ \bibinfo {author}
  {\bibfnamefont {T.}~\bibnamefont {Rizzo}},\ }\bibfield  {title} {\enquote
  {\bibinfo {title} {Finite-size corrections to disordered systems on
  {Erdös}-{Rényi} random graphs},}\ }\href {\doibase
  10.1103/PhysRevB.88.184201} {\bibfield  {journal} {\bibinfo  {journal}
  {Physical Review B}\ }\textbf {\bibinfo {volume} {88}},\ \bibinfo {pages}
  {184201} (\bibinfo {year} {2013})}\BibitemShut {NoStop}%
\bibitem [{\citenamefont {Ikeda}(2020{\natexlab{b}})}]{ikeda_jamming_crystals}%
  \BibitemOpen
  \bibfield  {author} {\bibinfo {author} {\bibfnamefont {Harukuni}\
  \bibnamefont {Ikeda}},\ }\bibfield  {title} {\enquote {\bibinfo {title}
  {Jamming and replica symmetry breaking of weakly disordered crystals},}\
  }\href {\doibase 10.1103/PhysRevResearch.2.033220} {\bibfield  {journal}
  {\bibinfo  {journal} {Physical Review Research}\ }\textbf {\bibinfo {volume}
  {2}},\ \bibinfo {pages} {033220} (\bibinfo {year}
  {2020}{\natexlab{b}})}\BibitemShut {NoStop}%
\bibitem [{Note1()}]{Note1}%
  \BibitemOpen
  \bibinfo {note} {Duke Digital Repository: \protect \url {
  https://doi.org/10.7924/r4833vm1m}}\BibitemShut {NoStop}%
\end{thebibliography}%

\end{document}